\documentclass{aa} 
\usepackage{graphicx}
\usepackage{deluxetable}
\usepackage{txfonts}
\usepackage{natbib}
\newcommand{\hii} {H\,{\sc ii}}
\newcommand{\hd}[1] {HD\,#1}
\newcommand{\hr}[1] {HR\,#1}
\newcommand{\vsini} {$v$\,sin\,$i$}
\newcommand{\Teff} {T$_{\rm eff}$}
\newcommand{\grav} {log\,{\em g}}
\newcommand{\kms} {km\,s$^{-1}$}
\newcommand{\micro} {$\xi_{\rm t}$}
\newcommand{\nlte} {{\sc nlte}}
\newcommand{\lte} {{\sc lte}}

\newcommand{\fastwind} {{\sc fastwind}}
\newcommand{\ioni}[2]{{#1\,\sc{#2}}}
\newcommand{\abun}[1]{{$\epsilon_{\rm #1}$}}
\newcommand{\fies}{{\sc fies}@{\sc not}}
\newcommand{\tsq}{t$^2$}
\begin{document}
\title{The chemical composition of the Orion star forming region}
\titlerunning{O and Si abundances in B-type stars in Ori\,OB1}
\subtitle{I. Homogeneity of O and Si abundances in B-type stars
\thanks{Based on observations made with the Nordic Optical Telescope, operated
jointly on the island of La Palma by Denmark, Finland, Iceland,
Norway, and Sweden, in the Spanish Observatorio del Roque de los
Muchachos of the Instituto de Astrofisica de Canarias.}}
\author{S. Sim\'on-D\'iaz\inst{1,2}}
\offprints{S. Sim\'on-D\'{\i}az, \email{ssimon@iac.es}}
   \institute{Instituto de Astrof\'isica de Canarias, E38200 La Laguna, Tenerife, Spain.              
             \and
             Departamento de Astrof\'isica, Universidad de La Laguna, E-38205 La Laguna, Tenerife, Spain.
             }
\date{Received; accepted}

% \abstract{}{}{}{}{} 
% 5 {} token are mandatory
\abstract
% context heading (optional)
% {} leave it empty if necessary  
{Recent accurate abundance analyses of B-type main sequence stars in the solar 
vicinity has shown that abundances derived from these stellar objects are more 
homogeneous and metal-rich than previously thought.} 
% aims heading (mandatory)
{We investigate whether the inhomogeneity of abundances previously found in 
B-type stars in the Ori\,OB1 association is real (hence a signature of 
enrichment of the newly formed stars in an induced star formation scenario) 
or a consequence of intrinsic errors induced by the use of photometric indices 
to establish the stellar parameters prior to the abundance analysis.}  
% methods heading (mandatory)
{We obtained a new (improved) spectroscopic data\,set comprising 13 B-type stars 
in the various Ori\,OB1 associations, and performed a detailed, self-consistent 
spectroscopic abundance analysis by means of the modern stellar atmosphere code 
\fastwind.} 
% results heading (mandatory)
{We detect systematic errors in the stellar parameters determined previously 
which affect the derived abundances. Once these errors are accounted for, we 
find a high degree of homogeneity in the O and Si abundances for stars in the 
four Ori\,OB1 subgroups. The derived abundances are in very good agreement 
with recent determinations in other B-type stars in the solar vicinity. We 
also compare our results with those obtained for the Sun during the epoch 
of the ``solar crisis", and the Orion nebula.} 
% conclusions heading (optional), leave it empty if necessary 
{}
\maketitle
%
%=======================================================================
\section{Introduction}\label{introduction}
%=======================================================================
For many years, our knowledge about the chemical composition of early-B main 
sequence stars in the solar vicinity has been characterized by two main results: 
(i) the derived abundances seemed to be highly inhomogeneous (with a dispersion 
of up to 0.5 dex), and (ii) the mean values indicated lower abundances than the 
standard \citep{Gre98} set of solar abundances \citep[see reviews by][]
{Her04, Mor08}. These results were not very encouraging, because the inhomogeneity 
of stellar abundances contradicted with the homogeneity in oxygen 
abundance found from studying the local diffuse interstellar medium 
\citep[e.g.][]{Mey98, Car06}. On the other hand, 
chemical evolution models of the Galaxy \citep[e.g.][]{Chi03, Car05} predict 
a small enrichment of the ISM in metals during the lifetime of the Sun (i.e. 
because they are younger than the Sun, nearby OB-type stars are expected to be slightly 
metal-rich).

Some recent results have began to change this situation. The solar oxygen 
abundance traditionally considered as a cosmic abundance reference \citep{Gre98} 
was reviewed by \cite{Asp04}, who obtained log\,(O/H)=8.66 dex, 0.17 dex lower 
than the standard value. This was the beginning of what has been called the
epoch of the``solar crisis": between 2004 and 2008, several studies by different 
authors \citep{Ayr06, Soc07, All08, Caf08, Ayr08, Mel08, Cen08} presented solar 
oxygen abundances derived by means of different approaches. The calculated values 
range from 8.63 dex \citep{Soc07} to 8.86 dex \citep{Cen08}. The debate about its
actual value is still open. 

\cite{Prz08} have recently analyzed a representative sample of six unevolved early 
B-type stars in nearby OB associations and the field, and found a very narrow 
distribution of abundances, with mean values that are more metal-rich compared to previous 
works (e.g. log\,(O/H)=8.76 dex, a value that is within the range of solar 
abundances calculated during the ``solar  crisis"). These authors indicate the 
importance of 
properly determining the atmospheric parameters and using robust model 
atoms to avoid systematic errors in the abundance determination. \citep[See also]
[for a summary of the main sources of systematic errors affecting the abundance 
analyses of B-type stars.]{Nie09} The study by \citeauthor{Prz08} show that the
chemical inhomogeneity previously found for B-type stars in the solar vicinity
may be spurious and an artificial effect of those systematic errors.
It also reinforces the importance of self-consistent spectroscopic abundance 
analyses (i.e., the stellar parameters and the metal abundances are determined 
exclusively from spectroscopic diagnostics by using the same set of stellar 
atmosphere models). Less accurate photometric \Teff\ estimates must be handle
with care (or better avoided whenever possible!) in the abundance analysis
of B-type stars \citep[see also][]{Nie08}. 

The Orion complex, containing the Orion molecular cloud and the Orion\,OB1
(Ori\,OB1) association, is one of the most massive active star-forming regions in the 
1 kpc centered on the Sun. \cite{Bla64} divided Ori\,OB1 into four 
subgroups of stars --- namely Ia, Ib, Ic, and Id --- having different 
locations in the sky and ages. \cite{Bro94} derived mean ages of 
11.4$\pm$1.9, 1.7$\pm$1.1\footnote{\cite{Bri05} determined an age for Ori\,OB1b
$\sim$4\,--\,6 Myr by studying its low mass, young stellar population; this value
is more consistent with what would be expected from the presence of the evolved
blue supergiant $\epsilon$\,Ori in this subgroup.}, 4.6$\pm$2, and $<$1 Myr for subgroups Ia to Id,
respectively. The youngest subgroup Ori\,OB1 Id is associated with the 
Orion nebula (M\,42), the most studied \hii\ region and the closest
ionized nebula to the Sun in which a high accuracy abundance analysis can 
be performed.

The correlation between the ages of the stellar subgroups, their location, 
and the large scale structures in the interstellar medium around 
Orion\,OB1 have been interpreted as features of sequential star formation 
and Type-II supernovae \citep{Rey79, Cow79, Bro94}. \cite{Cun92, Cun94} obtained 
C, N, O, Si, and Fe abundances of 18 B-type main sequence stars from the 
four subgroups comprising the Ori\,OB1 association. They found a range
in oxygen abundances\footnote{Other authors \citep{Gie92, Kil92, 
Gum98} obtained a similar range in oxygen abundances from the analysis 
of smaller samples of B-type stars in Orion\,OB1.} of $\sim$0.4 dex, with 
the highest values corresponding to the stars in the youngest (Id and 
some Ic) subgroups. 
In this case, the inhomogeneity in stellar abundances (mainly oxygen
and silicon) seemed to be real and coherent with a scenario of induced star 
formation in which the new generation of stars are formed from interstellar 
material contaminated by Type-II supernovae ejecta. 

The study by \citeauthor{Cun92} was based on a photometric estimation of 
\Teff\ and the fitting of the H$\gamma$ line computed from \cite{Kur79} \lte\ 
model atmospheres to the observed one to derive \grav. In a more recent work, 
\cite{Sim06} used one of the new generation of \nlte, line blanketed, model 
atmosphere codes \citep[\fastwind,][]{San97, Pul05} and a self-consistent 
spectroscopic approach to derive the stellar parameters and oxygen abundances 
for the three B0.5\,V stars in Ori\,OB1d (the youngest and, supposedly, more 
metal-rich subgroup). The resulting stellar parameters were somewhat different and, 
more important, the derived abundances were systematically lower than the 
previous values by \cite{Cun94} by $\sim$\,0.2--0.3 dex.

This result motivated us to review the chemical composition of the other 
B-type stars in Ori\,OB1 to investigate whether the inhomogeneity of 
abundances previously found is real or a consequence of intrinsic errors 
induced by the use of photometric indices to establish the stellar parameters 
prior to the abundance analysis.
To this aim, we obtained a totally new (improved) observational data\,set and
performed a self-consistent abundance analysis of 13 of the stars considered
by \cite{Cun92, Cun94}. 

We used \fastwind\ to derive the stellar parameters, oxygen, and 
silicon abundances. The observational data\,set is described in 
Sect.\,\ref{observa}. The whole spectroscopic analysis is presented in
Sect.\,\ref{analysis}. Then, we compare our stellar parameters and abundance
with results from previous works (Sect.\,\ref{compare}). The homogeneity 
of stellar abundances in Ori\,OB1 and its comparison with other B-type stars 
in the solar vicinity determinations, the Sun, and the Orion nebula is 
discussed in Sect.\,\ref{discuss}. The main conclusions of this work are
summarized in Sect.\,\ref{conclude}.

%=======================================================================
\section{Observational data\,set}\label{observa}
%=======================================================================
%%%%%%%%%%%%%%%%%%%%%%%%%%%%%%%%%%%%%%%%%%%%%%%%%%%%%%%%%%%%%%%%%%%%%%%%%%%%
\begin{figure*}[!t]
\centering
\includegraphics[angle=90,scale=0.62]{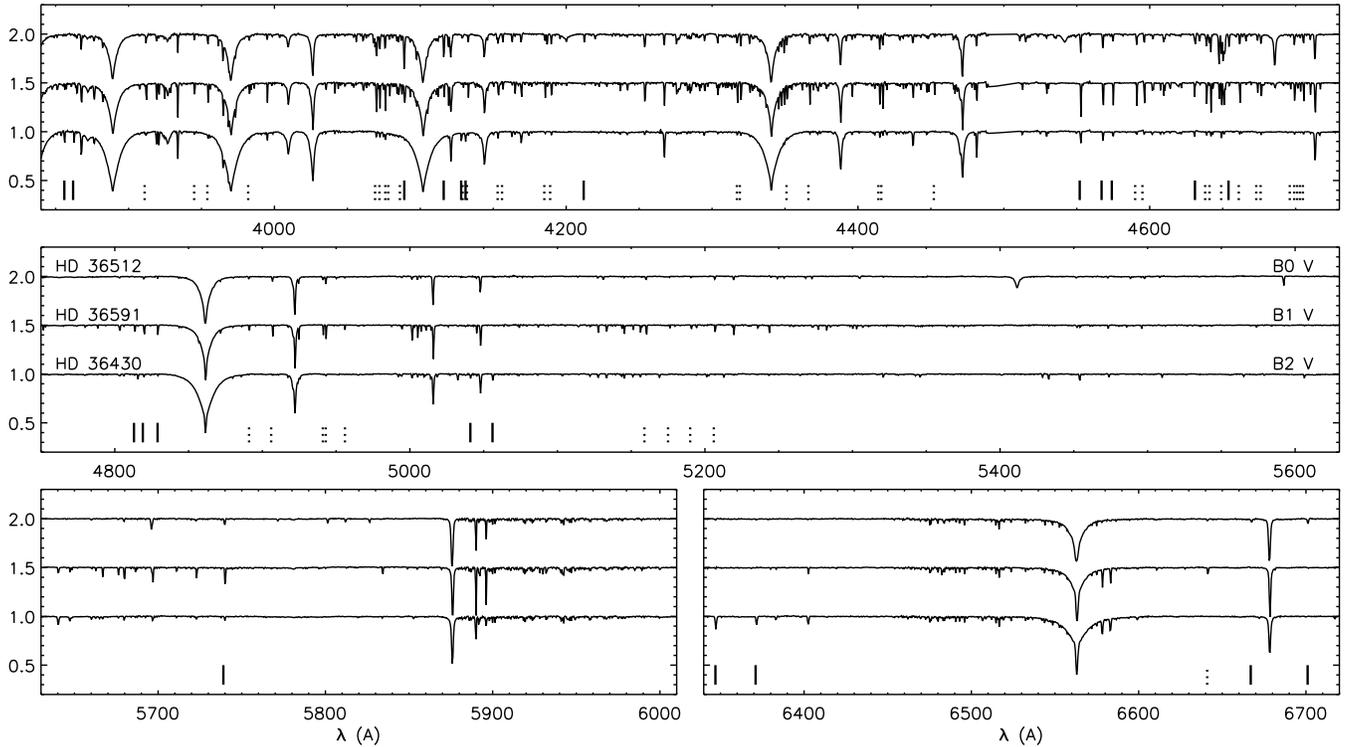}
\caption{Example of \fies\ spectra for three of the observed 
stars. The complete atlas is available in the electronic
version of the paper (Fig.\,\ref{fig1e}). The \ioni{Si}{ii-iv} and \ioni{O}{ii} lines used 
for the abundance analysis are indicated as solid and dashed vertical 
lines, respectively. \label{fig1}}
\end{figure*}
%%%%%%%%%%%%%%%%%%%%%%%%%%%%%%%%%%%%%%%%%%%%%%%%%%%%%%%%%%%%%%%%%%%%%%%%%%%%

The observations used here were carried out with the FIES cross-dispersed, 
high-resolution echelle spectrograph attached to the NOT2.5m telescope at 
El Roque de los Muchachos observatory on La Palma (Islas Canarias, Spain) 
on 5-8 November 2008. The medium-resolution mode (R=46000, $\delta\lambda$=0.03 
\AA/pix) was selected, and the entire spectral range 3700-7300 \AA\ was 
covered without gaps in a single fixed setting. A sample of 14 stellar 
candidates was observed, after being selected from the list of stars in the Ori\,OB1 association 
analyzed by \cite{Cun92, Cun94}. These are early-B type, non-evolved, 
stellar objects (B0\,V-B2\,V) with low projected rotational 
velocities (\vsini\,$\leq$\,60 \kms). 
The signal-to-noise ratio achieved for all the spectra was 
always above 250. The list of observed stars is presented in Table\,\ref{table1}. 
Apart from the new set of stars, we re-observed \hd{37020} 
and \hd{37042}, two of the three B-type stars\footnote{\hd{37023} 
was also re-observed, but it was not included in the analysis because we found 
in the new spectrum that the star is is actually a binary system (SB2), 
hence not optimal for an abundance analysis.} analyzed in \cite{Sim06}. 
 
The spectra were reduced with the FIEStool\footnote{http://www.not.iac.es/instruments/fies/fiestool/FIEStool.html} 
software in advanced mode. The FIEStool pipeline provided wavelength calibrated, 
blaze-corrected, order-merged spectra of high quality. These spectra were then 
normalized with our own developed {\sc idl} programs. An example of the 
\fies\ spectra for three of the observed stars is presented in
Fig \ref{fig1}, where the \ioni{Si}{ii-iv} and \ioni{O}{ii} lines used 
for the abundance analysis are also indicated.

%=======================================================================
\section{Spectroscopic analysis}\label{analysis}
%=======================================================================

The analyses were performed following a self-consistent spectroscopic 
approach with the spherically extended, \nlte, line-blanketed stellar 
atmosphere code \fastwind\ \citep{San97, Pul05}. Basically, the 
stellar parameters were derived by comparing the observed H Balmer line 
profiles and the ratio of \ioni{Si}{iii-iv} and/or \ioni{Si}{ii-iii} 
line equivalent widths with the output from a grid of \fastwind\ models.
Whenever possible, the \ioni{He}{i-ii} ionizing equilibrium was also
considered. Then, the same grid of models was used to derive the stellar 
abundances by means of the curve-of-growth method.

%=======================================================================
\subsection{Projected rotational velocities}
%=======================================================================

% -----------------------------------------------------------------------
% --------------------           TABLE 1         ------------------------
% -----------------------------------------------------------------------
%
\begin{table}[!t]
\begin{center}
\caption{List of B-type stars from Ori\,OB1 considered in this study, 
ordered by spectral type. 
\label{table1}}
\begin{tabular}{cccccc}
\tableline
\tableline
\noalign{\smallskip}
HD  & HR  & Name & SpT & V & Subgroup \\
\tableline
\noalign{\smallskip}
\hd{36512} & \hr{1855} & $\nu$\,Ori         & B0\,V   & 4.62 &  Ic \\
\hd{37020} & \hr{1893} & $\theta^1$\,Ori\,A & B0.5\,V & 6.71 &  Id \\
\hd{36960} & \hr{1887} &                    & B0.5\,V & 4.78 &  Ic \\
\hd{37042} &           & $\theta^2$\,Ori\,B & B0.7\,V & 6.02 &  Id \\
\hd{36591} & \hr{1861} &                    & B1\,V   & 5.34 &  Ib \\
\hd{36959} & \hr{1886} &                    & B1\,V   & 5.67 &  Ic \\
\hd{35299} & \hr{1781} &                    & B1.5\,V & 5.69 &  Ia \\
\hd{37744} & \hr{1950} &                    & B1.5\,V & 6.22 &  Ib \\
\hd{36285} & \hr{1840} &                    & B2\,V   & 6.33 &  Ic \\
\hd{36629} &           &                    & B2\,V   & 7.65 &  Ic \\
\hd{35039} & \hr{1765} &                    & B2\,V   & 4.72 &  Ia \\
\hd{36430} & \hr{1848} &                    & B2\,V   & 6.23 &  Ic \\
\hd{35912} & \hr{1820} &                    & B2\,V   & 6.41 &  Ia \\   
\noalign{\smallskip}
\tableline
\end{tabular}
\end{center}
\end{table}
% -----------------------------------------------------------------------
% -----------------------------------------------------------------------

The projected rotational velocities (\vsini) were obtained by applying
the Fourier method (\citeauthor{Gra76}, \citeyear{Gra76}; see also 
\citeauthor{Sim07}, \citeyear{Sim07}, for its application to OB-type stars) 
to the \ioni{Si}{iii}\,$\lambda$4552 line. The 0.03 \AA/pix 
resolution of the FIES spectra implies that the lowest \vsini\ that could be 
detected is 2-3 \kms; however, for those cases with \vsini\,$<$\,10-15 \kms, 
identifying of the first zero of the Fourier transform was difficult 
due to the effect of the noise (an maybe the microturbulence).

In many cases, an additional extra Gaussian-type broadening {\bf ($\Theta_{g}$)} was needed to properly
fit the line profile for the derived value of \vsini. This extra-broadening
account for the microturbulence (and maybe the macroturbulence), also affecting
the shape of the line. The corresponding derived values are summarized in Table 
\ref{table2}.

%=======================================================================
\subsection{Measurement of equivalent widths and identification of problematic lines}
%=======================================================================

% -----------------------------------------------------------------------
% --------------------           TABLE 2         ------------------------
% -----------------------------------------------------------------------
%
\begin{table}[!t]
\begin{center}
\caption{Projected rotational velocities derived for the studied stars.
\label{table2}}
\begin{tabular}{cccccc}
\noalign{\smallskip}
\tableline
\tableline
\noalign{\smallskip}
HD  & \vsini$^a$  & $\Theta_{g}^a$ & HD & \vsini & $\Theta_{g}$ \\
\tableline
\noalign{\smallskip}
\hd{36512} &  15   & 30 &  \hd{37744} &   37  & 00\\ 
\hd{37020} &  55   & 00 &  \hd{36285} & $<$10 & 20\\ 
\hd{36960} &  23   & 40 &  \hd{36629} & $<$10 & 15\\ 
\hd{37042} &  31   & 00 &  \hd{35039} & $<$12 & 15\\   
\hd{36591} &  08   & 19 &  \hd{36430} & $<$15 & 25\\  
\hd{36959} &  12   & 12 &  \hd{35912} & $<$12 & 20\\ 
\hd{35299} & $<$08 & 12 &  \\ 
\noalign{\smallskip}
\tableline
\end{tabular}
\tablenotetext{a}{\footnotesize Both quantities in \kms.}
\end{center}
\end{table}
% -----------------------------------------------------------------------
% -----------------------------------------------------------------------

The strategy we followed in our analyses is based on the equivalent widths (EW) 
of metal lines. Therefore, proper identification of the lines of interest,
along with an accurate measurement of their $EW$s (also including the associated 
uncertainties), is a very important step.
We have developed IDL routines automatically identifying metal lines in the spectra, 
measuring the $EW$s and their uncertainties, and detecting the possibility that other 
lines affect these measurements. To this aim we compiled a list of C, N, O, Si, Mg, S,
Ne, and Ar lines, extracted from the atomic\footnote{http://www.pa.uky.edu/$\sim$peter/newpage/} 
line list v2.05.
For a given line, the program performs a multi-Gaussian fit of the observed line 
profile accounting for all the lines expected to be present in a certain spectral range
($\lambda_0$$\pm$2\,max[\vsini$\lambda_0$/c, 0.5$\lambda_0$/R]) around the wavelength 
indicated in the line list. The uncertainty in the $EW$ measurement is obtained by assuming 
the location of the local continuum at $\pm$1/SNR and, in those cases in which the line 
is isolated, comparing the value obtained by means of the Gaussian fitting with the value 
derived by integrating the line.

The high quality of our spectra allowed us to identify and measure the $EW$s of up to 27 
\ioni{Si}{ii-iv} lines and 47 \ioni{O}{ii} lines. Some of the lines were labeled as 
problematic because of the presence of one or more lines from other elements (e.g. the 
\ioni{O}{ii}\,$\lambda$4673.73 line may be affected by the \ioni{C}{iii}\,$\lambda$4673.95 
in same cases; similarly occurs for the 
\ioni{O}{ii}\,$\lambda$4641.81 line, coincident with the \ioni{N}{iii}\,$\lambda$4641.85 line).
These lines were treated with special care in the abundance analysis, since they may be giving
wrong values for the abundance.

Generally, the Gaussian fit provides reliable results for the $EW$s; however,
in those cases in which the \vsini\ of the star is above 30-35 \kms, the use of a Gaussian
to fit the profile may result in an under or overestimation of the $EW$ (in a few percent), 
depending on the line strength. For those cases, the $EW$ resulting from the integration of 
the observed line profile was preferred.

%=======================================================================
\subsection{A grid of HHeOSi \fastwind\ models}
%=======================================================================

For this study, we constructed a grid of \fastwind\ models with \Teff\ and 
\grav\ ranging from 17 to 36 kK (1 kK steps) and 3.7 to 4.3 dex (0.1 dex 
steps). As the studied stars are not expected to be evolved, the He abundance 
was fixed to 0.09 dex. In addition, since \fastwind\ is a spherically extended 
code, the radius and other wind parameters need to be indicated, and are grouped 
in the Q-parameter. We fixed this parameter to log\,Q=\,--\,15 as a representative value
for which the wind effect over the optical spectrum is practically negligible.
(H$\alpha$ and \ioni{He}{ii}$\lambda$4686 show no sign of wind contamination.) The 
metallicity was assumed to be solar \citep[following the set of abundances by]
[]{Gre98}.

For each pair of stellar parameters, a sub\,grid of models varying the microturbulence 
(\micro\,=\,1, 3, 5, 7, 9 \kms), the Si abundance ($\epsilon_{\rm Si}$\,=\,--5.10, 
--4.80, --4.50, --4.20 dex), and the O abundance ($\epsilon_{\rm O}$\,=\,--4.00, 
--3.65, --3.30, --2.95 dex) was calculated. The O and Si atomic models used for the 
grid came mainly from \cite{Bec88, Bec90}. However, two updates were considered: (a) a extended
\ioni{Si}{ii} model atom \cite[see][]{Tru04}, and (b) the most recent log\,gf 
values indicated in the atomic line list v2.05 for the formal solution calculations. 

The final grid consists of 20x7x4 (=560) models and a total of 2800 formal solutions
(5 microturbulence values per model). It includes line profiles and $EW$s
for \ioni{H}{}, \ioni{He}{i-ii}, \ioni{Si}{ii-iv} and \ioni{O}{ii} lines, along with
the spectral energy distribution for each set of stellar parameters.

%=======================================================================
\subsection{Determining stellar parameters}\label{parameters}
%=======================================================================

% -----------------------------------------------------------------------
% --------------------           TABLE 3         ------------------------
% -----------------------------------------------------------------------
%
\begin{table*}[!t]
\tabcolsep=4.pt
\begin{center}
\caption{Final results of the HSi analysis: stellar parameters and Si abundances.
Detailed results (line-by-line) of the Si abundance analysis are presented in Tables
\ref{res1} to \ref{res13} in the electronic version of the paper. All \Teff\ values 
correspond to the \grav\ indicated in the 8th column.
\label{table3}}
\begin{tabular}{ccccccccccccc}
\noalign{\smallskip}
\tableline
\tableline
\noalign{\smallskip}
 & & {\scriptsize ($EW_{\rm 4}$/$EW_{\rm 3}$)$^a$} 
 & {\scriptsize ($EW_{\rm 2A}$/$EW_{\rm 3}$)}
 & {\scriptsize ($EW_{\rm 2B}$/$EW_{\rm 3}$)} 
 &  & \multicolumn{2}{c}{Adopted} & & & \Teff$\pm$500\,K\\
\cline{7-8}
Target     &   SpT   & 
\Teff\ -- \ioni{Si}{iv-iii} & 
\Teff\ -- \ioni{Si}{ii-iii} & 
\Teff\ -- \ioni{Si}{ii-iii} & 
\Teff\ -- \ioni{He}{i-ii} & \Teff\ & \grav\ & \micro(Si) & \abun{Si}$\pm$$\Delta$\abun{Si}[$\sigma$, \micro] & $\Delta$\abun{Si} %%@
(\ion{Si}{iii}) \\
\tableline
\noalign{\smallskip}
   & & {\scriptsize (1.42$\pm$0.06)} &  &  & &  & & \\
\hd{36512}  & B0\,V   & 33700$\pm$200 &     ...       & ... & 34000$\pm$500 & {\bf 33700} & {\bf 4.2} &  4.3$\pm$0.7 &  {\bf %%@
7.49}$\pm$\,[0.07, 0.05] & $\pm$0.12 \\
   & & {\scriptsize (0.69$\pm$0.09)} &  &  & &  & & \\
\hd{37020}  & B0.5\,V & 30500$\pm$600 &    ...       & ... & 30000$\pm$500 & {\bf 30500} & {\bf 4.2} &  0.5$\pm$0.5 &  {\bf %%@
7.47}$\pm$\,[0.10, 0.04] & $\pm$0.06 \\
   & & {\scriptsize (0.60$\pm$0.04)} &  &  & &  & & \\
\hd{36960}  & B0.5\,V & 28900$\pm$300 &     ...       & ... & 29000$\pm$500 & {\bf 28900} & {\bf 3.9} &  5.4$\pm$0.6 &  {\bf %%@
7.53}$\pm$\,[0.02, 0.06] & $\pm$0.06 \\
   & & {\scriptsize (0.64$\pm$0.07)} &  &  & &  & & \\
\hd{37042}  & B0.7\,V & 29700$\pm$400 &     ...       & ... & 29500$\pm$500 & {\bf 29700} & {\bf 4.2} &  1.4$\pm$0.3 &  {\bf %%@
7.55}$\pm$\,[0.03, 0.04] & $\pm$0.05 \\
   & & {\scriptsize (0.39$\pm$0.01)} &  &  & &  & & \\
\hd{36591}  & B1\,V   & 27200$\pm$100 &     ...       & ... & ... & {\bf 27200} & {\bf 4.1} &  1.3$\pm$0.3 &  {\bf 7.53}$\pm$\,[0.06, 0.03]  %%@
& $\pm$0.01 \\   
   & & {\scriptsize (0.25$\pm$0.02)} & {\scriptsize (0.12$\pm$0.02)} &                               & &  & & \\
\hd{36959}  & B1\,V   & 25900$\pm$300 & 25900$\pm$100 &   ...         & ... & {\bf 25900} & {\bf 4.2} &  0.0$\pm$1.0 &  {\bf %%@
7.50}$\pm$\,[0.05, 0.07]  & $\mp$0.02    \\
   & & {\scriptsize (0.16$\pm$0.03)} & {\scriptsize (0.12$\pm$0.01)} & {\scriptsize (0.12$\pm$0.04)} & &  & & \\
\hd{37744}  & B1.5\,V & 23900$\pm$600 & 25700$\pm$100 & 23600$\pm$600 & ... & {\bf 23800} & {\bf 4.1} & 0.5$\pm$0.5 & {\bf %%@
7.54}$\pm$\,[0.06, 0.04] &  $\mp$0.05 \\
   & & {\scriptsize (0.14$\pm$0.02)} & {\scriptsize (0.20$\pm$0.01)} & {\scriptsize (0.17$\pm$0.03)} & &  & & \\
\hd{35299}  & B1.5\,V & 23900$\pm$300 & 24900$\pm$300 & 23700$\pm$300 & ... & {\bf 23700} & {\bf 4.2} &	0.5$\pm$0.5	 & {\bf %%@
7.50}$\pm$\,[0.08, 0.02]  & $\mp$0.06 \\
   & &                               & {\scriptsize (0.26$\pm$0.01)} & {\scriptsize (0.41$\pm$0.03)} & &  & & \\
\hd{36285}  & B2\,V   &    ...        & 23900$\pm$200 & 20600$\pm$200 & ... & {\bf 20600} & {\bf 4.0} & 1.7$\pm$0.5 & {\bf %%@
7.49}$\pm$\,[0.06, 0.05]  & $\mp$0.11 \\
   & &                               & {\scriptsize (0.36$\pm$0.02)} & {\scriptsize (0.44$\pm$0.04)} & &  & & \\
\hd{35039}  & B2\,V   &    ...        & 22200$\pm$200 & 19900$\pm$200 & ... & {\bf 19800} & {\bf 3.7} & 3.3$\pm$1.0 & {\bf %%@
7.52}$\pm$\,[0.06, 0.08]  & $\mp$0.14 \\
   & &                               & {\scriptsize (0.39$\pm$0.02)} & {\scriptsize (0.53$\pm$0.02)} & &  & & \\
\hd{36629}  & B2\,V   &    ...        & 22800$\pm$300 & 20000$\pm$100 & ... & {\bf 20000} & {\bf 4.1} & 1.0$\pm$0.5 & {\bf %%@
7.54}$\pm$\,[0.04, 0.05] & $\mp$0.13 \\
   & &                               & {\scriptsize (0.65$\pm$0.06)} & {\scriptsize (0.96$\pm$0.09)} & &  & & \\
\hd{36430}  & B2\,V   &    ...        & 21000$\pm$300 & 18600$\pm$200 & ... & {\bf 18600} & {\bf 4.1} & 3.5$\pm$1.0 & {\bf %%@
7.47}$\pm$\,[0.08, 0.07]  & $\mp$0.13 \\
   & &                               & {\scriptsize (0.73$\pm$0.05)} & {\scriptsize (0.92$\pm$0.05)} & &  & & \\
\hd{35912}  & B2\,V   &    ...        & 20400$\pm$300 & 18500$\pm$150 & ... & {\bf 18500} & {\bf 4.0} & 3.2$\pm$0.5 & {\bf %%@
7.48}$\pm$\,[0.07, 0.04]  & $\mp$0.13 \\
\noalign{\smallskip}
\tableline
\end{tabular}
\tablenotetext{a}{\footnotesize $EW_{\rm 4}$: \ioni{Si}{iv}\,4116; $EW_{\rm 3}$: \ioni{Si}{iii}\,4552; 
$EW_{\rm 2A}$: \ioni{Si}{ii}\,4130; $EW_{\rm 2B}$: \ioni{Si}{ii}\,6371}  
\end{center}
\end{table*}
% -----------------------------------------------------------------------
% -----------------------------------------------------------------------

The use of the \ioni{Si}{iii-iv} and/or \ioni{Si}{ii-iii} ionization 
equilibrium, along with the H Balmer lines, for determining the 
stellar parameters of early B-type stars is a longstanding method described 
elsewhere \citep[see e.g.][]{Kil91, Urb05, Cro06, Mar08}. 
The ratios $EW$(\ioni{Si}{iv}\,$\lambda$4116)/$EW$(\ioni{Si}{iii}\,$\lambda$4552)
and/or $EW$(\ioni{Si}{ii}\,$\lambda$4128)/$EW$(\ioni{Si}{iii}\,$\lambda$4552),
depending on the temperature of the star have been traditionally used as \Teff\
indicators. This decision has been probably motivated by the fact that these 
are the strongest, unblended lines in the spectral range commonly observed 
for the stellar abundance analysis (i.e. $\sim$\,4000\,--\,5000 \AA).

We derived the stellar parameters, along with the Si abundance, for all 
stars in our sample using this methodology. We initially considered that these 
Si line ratios determine the effective temperatures 
(columns 3 and 4 in Table \ref{table3}). However, motivated by a couple of 
problems found when deriving the stellar parameters in the cooler objects 
(see discussion below), we decide to include the $EW$(\ioni{Si}{ii}\,$\lambda$6347)/$EW$(\ioni{Si}{iii}\,$\lambda$4552) 
ratio as a temperature indicator (column 5). The measured values are indicated in brackets in the corresponding
columns. As expected, the \ioni{Si}{iv}/\ioni{Si}{iii} ratio decreases when 
we move to later spectral types, and the \ioni{Si}{ii}/\ioni{Si}{iii} behaves 
in the opposite way. For three of the stars (\hd{36959}, \hd{35299}, and \hd{37744}), 
lines from the three ions are clearly and simultaneously present in the spectra. 

Four parameters need to be determined at the same time in an iterative way: 
\Teff, \grav, \micro(Si), 
and \abun{Si}. First, we use the Si line ratios indicated above and the wings of 
the H Balmer lines (fixing \micro(Si), and \abun{Si}) to obtain an initial guess for 
\Teff\ and \grav. Then we apply the curve of growth method to a proper set of 
\ioni{Si}{ii-iv} lines to iteratively obtain final values for the four parameters 
(a detailed description of the used lines and the results of the Si abundance analysis 
is presented below). Normally, the final values of \Teff\ and \grav\ are 
quite close to the initial values, since the mentioned line ratios are only slightly 
dependent on \abun{Si} and \micro.

\subsubsection{Hotter objects (\Teff $\ge$ 27000 K)}
In the hotter objects, \ion{Si}{ii} lines are not present in the spectra. The 
$EW$(\ioni{Si}{iv}\,$\lambda$4116)/$EW$(\ioni{Si}{iii}\,$\lambda$4552) ratio
is thus used to obtain the initial guess values of \Teff\ for each of the \grav\
values considered in the grid. Then, the full set of reliable \ion{Si}{iii-iv} 
lines, along with the H Balmer line profiles, is used for fine determination 
of the four parameters indicated above (see an example in the upper panel of 
Fig. \ref{fig2}, and in Fig. \ref{fig3}). The uncertainties in the $EW$ measurements of the Si 
lines have been taken into account for establishing the uncertainties associated 
with the derived stellar parameters. Generally, the \Teff\ and \grav\ can be 
determined with an accuracy better than 500--600\,K and 0.1\,dex, respectively. The final 
results of the analysis are presented in Table\,\ref{table3}. 

For main sequence stars with \Teff\,$\gtrsim$\,28000 K the \ioni{He}{i-ii} ionization 
equilibrium can also be used \citep[see e.g.][]{Her92, Sim06}. (Below this temperature,
\ioni{He}{ii} lines are too faint or not present in the spectrum.) 
We could determine the stellar parameters in this way for four of the stars in our sample.
An example of this type of analysis has already been presented in \cite{Sim06}, 
so we only present here the corresponding results (see column 6 in 
Table \ref{table3}). In general, there is good agreement between the \Teff\ determined through the 
\ioni{Si}{iii-iv} and the \ioni{He}{i-ii} ionization balance (with differences in
\Teff\ not larger than $\sim$500 K.

\subsubsection{Cooler objects}

For the cooler objects, the \ioni{Si}{ii}/\ioni{Si}{iii} must be used. We initially
considered the $EW$(\ioni{Si}{ii}\,$\lambda$4130)/$EW$(\ioni{Si}{iii}\,$\lambda$4552)
ratio for obtaining a first estimation of the stellar parameters; however, we found two
facts that warned us of a possible problem with the \ioni{Si}{ii}\,$\lambda$4130
(or the \ioni{Si}{ii}\,$\lambda$4128) line. First, it was not possible to properly
fit the H Balmer lines for any of the (\Teff, \grav)-pairs indicated by
the \ioni{Si}{ii}\,$\lambda$4130/\ioni{Si}{iii}\,$\lambda$4552 line ratio, because the core 
of the \ioni{H}{} synthetic lines were somewhat narrower than the observed ones (even
if high values of \grav\ are considered). For
these temperatures, the cores of the H Balmer lines begin to be sensitive to changes
in \Teff. For all the cases studied, these lines require somewhat lower effective
temperatures to properly fit their cores.
Second, the \abun{Si}-EW diagnostic diagrams show two sets of \ioni{Si}{ii} 
lines giving different abundances by $\sim$0.2 dex (see lower panel in Fig.\,\ref{fig2}). 
Curiously, 
each subset of lines corresponds to transitions with very different energy levels 
(\ioni{Si}{ii}\,$\lambda$$\lambda$4128, 4130, 5041, 5056, in one hand, coming from higher 
energy levels; \ioni{Si}{ii}\,$\lambda$$\lambda$6347, 6371, 3856, 3862, on the other, 
comming from lower energy levels; we suggest the reader have a Grotrian diagram in hand). 
This different behavior of the various lines warned us of possible problems 
with the \ioni{Si}{ii} atomic model and of using \ioni{Si}{ii}\,$\lambda$4128 line 
alone to establish the stellar temperature.

Columns 4 and 5 in Table \ref{table3} show the different effective temperatures 
obtained depending on the \ioni{Si}{ii} line that is used 
(\ioni{Si}{ii}\,$\lambda$4128, \ioni{Si}{ii}\,$\lambda$6371). Temperatures 
given by the $EW$(\ioni{Si}{ii}\,$\lambda$6371)/$EW$(\ioni{Si}{iii}\,$\lambda$4552) 
ratio are systematically lower by up to $\sim$2000\,--\,3000 K. In addition, 
a proper fit of the cores of the H Balmer lines can be achieved 
with these new values of the temperature. 

\subsubsection{\hd{36595}, \hd{37744}, and \hd{35299}}

These three stars show \ioni{Si}{ii-iii-iv} lines in their spectra, so can 
contribute with new decisive clues to the problem mentioned above. In this
case we count with three \Teff\ indicators. As can be noticed from inspection
of Table \ref{table3}, the \Teff\ indicated by the 
$EW$(\ioni{Si}{ii}\,$\lambda$6371)/$EW$(\ioni{Si}{iii}\,$\lambda$4552) ratio
results in better agreement than the other \ioni{Si}{ii}/\ioni{Si}{iii} ratio
for two of the three cases. For the hotter object (\hd{36959}), the 
\ioni{Si}{ii}\,$\lambda$6371 is too faint to be measured. Although the
\Teff\ indicated by the 
$EW$(\ioni{Si}{ii}\,$\lambda$4130)/$EW$(\ioni{Si}{iii}\,$\lambda$4552) ratio 
agrees with the \ioni{Si}{iv}/\ioni{Si}{iii} diagnostic in this object, this is 
not the case for the other objects.

\begin{figure}[!t]
\centering
\includegraphics[angle=0,scale=0.43]{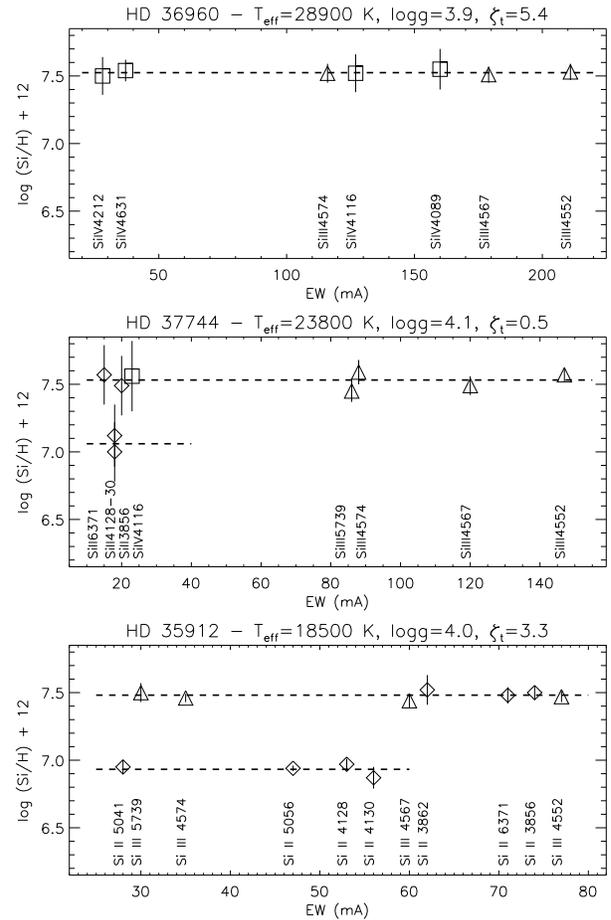}
\caption{Silicon abundance vs. $EW$ diagnostic diagrams for 
three of the analyzed stars (with representative effective temperatures).\label{fig2}}
\end{figure}

In Fig. \ref{fig2} (middle panel) we show the \abun{Si}-EW diagnostic diagram for 
\hd{37744}. The stellar parameters considered in this plot are those indicated
by the \ioni{Si}{iv}/\ioni{Si}{iii} ratio. The \ioni{Si}{ii}\,$\lambda$$\lambda$6371, 3856
lines fit the other \ioni{Si}{iii-iv} lines perfectly, but not the 
\ioni{Si}{ii}\,$\lambda$$\lambda$4128, 4130 lines.

\subsubsection{Concluding}

%%%%%%%%%%%%%%%%%%%%%%%%%%%%%%%%%%%%%%%%%%%%%%%%%%%%%%%%%%%%%%%%%%%%%%%%%%%%
%%%%%%%%%%%%%%%%%%%%%%%%%%%%%%%%%%%%%%%%%%%%%%%%%%%%%%%%%%%%%%%%%%%%%%%%%%%%

%%%%%%%%%%%%%%%%%%%%%%%%%%%%%%%%%%%%%%%%%%%%%%%%%%%%%%%%%%%%%%%%%%%%%%%%%%%%
\begin{figure}[!t]
\centering
\includegraphics[angle=0,scale=0.43]{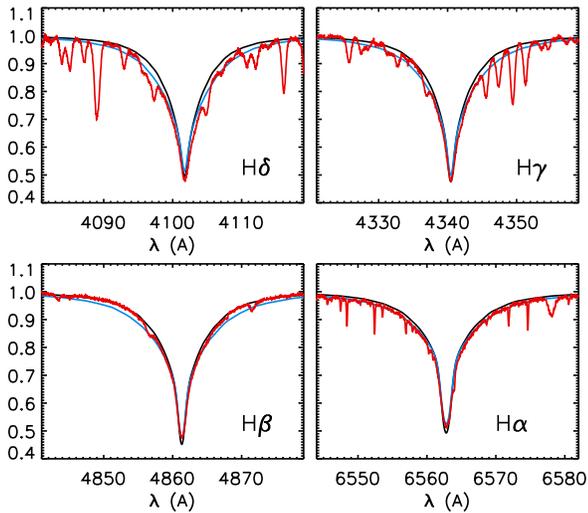}
\caption{Example of fitting the H Balmer lines to determine \grav. 
Two \fastwind\ models are compared with the observed spectrum (red line)
of \hd{36960}: in black, model with the stellar parameters resulting from the
\fastwind\ HSi analysis (28900, 3.9); in blue, model with the stellar parameters
derived by \cite{Cun92}, for comparison (28900, 4.3). \label{fig3}}
\end{figure}
%%%%%%%%%%%%%%%%%%%%%%%%%%%%%%%%%%%%%%%%%%%%%%%%%%%%%%%%%%%%%%%%%%%%%%%%%%%%

We found some indications of a problem in the \ioni{Si}{ii} model atom
that led to bad modeling of the \ioni{Si}{ii}\,$\lambda$$\lambda$4128, 4130, 5041, 5056 lines.
Fortunately, our observations also include other \ioni{Si}{ii} lines that seem to 
behave properly. A review of the \ioni{Si}{ii} model atom is needed (and could be 
tested with the type of detailed analysis presented here), but in the meantime, several
arguments allow us to trust\footnote{This hypothesis is also supported by a comparison of
\Teff\ obtained for the cooler objects in our sample from other spectroscopic diagnostic 
based on \ioni{O}{ii/i} and \ioni{C}{iii/ii} ratios of lines (F. Nieva, private 
communication).} the \ioni{Si}{ii}\,$\lambda$$\lambda$6347, 6371, 3856, 3862 set 
of lines for the stellar parameter determination
and for abundance analysis of the cooler objects in our sample: (a) the coherence of results
between \ioni{Si}{iv-iii} and \ioni{Si}{iii-ii} in terms of \Teff; (b) the good fit of the
H Balmer lines for the cooler stars when the lower \Teff\ is considered; (c) the coherence
of results in Si and O abundances that is obtained for all the stars in our sample when this solution 
is adopted (see below). 

\subsubsection{Uncertainties}

Columns 3 and 5 indicate the uncertainties in \Teff\ obtained from considering 
errors in the $EW$s measurement of \ioni{Si}{ii-iv} lines. These uncertainties, 
obtained by assuming a fixed gravity are $\sim$\,100-600\,K. Gravity can be normally established 
with an accuracy better 0.1 dex. Uncertainties from both quantities are correlated, e.g. a positive
variation of \grav\ of 0.1 dex needs to be compensated by an increase of \Teff\ $\sim$100-300\,K
to guarantee again the Si ionization balance. Although in many cases the formal errors in 
temperature obtained from the propagation of $EW$s errors are smaller than 500\,K, our 
experience warns us to be conservative (since other parameters can also slightly affect the 
derived temperatures, such as the considered microturbulence). Therefore, we adopt 500\,K and 
0.1 dex as characteristic uncertainties in \Teff\ and \grav, respectively, from our analyses.
  
%=======================================================================
\subsection{Silicon and oxygen stellar abundances}
%=======================================================================

We applied the curve-of-growth method to derive Si and O abundances. 
This method considers a grid of models for a given set of stellar parameters
(\Teff\ and \grav) in which the microturbulence and the abundance of the
element to be studied are varied. An abundance is obtained for the various 
values of microturbulence for each of the considered lines (given the measured 
$EW$ of the lines in the observed spectrum). Then, the final abundance is 
given by that microturbulence that results in all lines giving the same 
abundance. More details on this method can be found in e.g. \cite{Kil92} or 
\cite{Sim06}. 

Our preference for this methodology in performing the abundance analysis is
that we find the curve-of-growth method very powerful for identifying problematic 
lines (as shown e.g. in the previous section), which can affect the final 
abundance determination, and provide precise estimations of the uncertainties 
associated with the dispersion of line-to-line abundances, microturbulence, 
and the stellar parameters.

%=======================================================================
\subsubsection{Silicon abundances}
%=======================================================================

%%%%%%%%%%%%%%%%%%%%%%%%%%%%%%%%%%%%%%%%%%%%%%%%%%%%%%%%%%%%%%%%%%%%%%%%%%%%
\begin{figure}[!t]
\centering
\includegraphics[angle=0,scale=0.45]{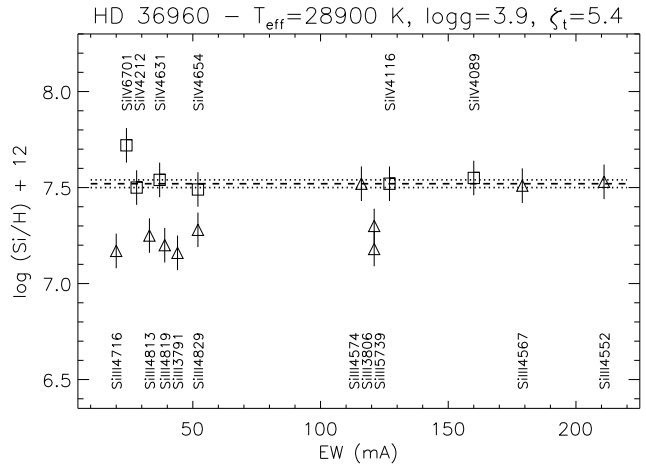}
\caption{Silicon abundance vs. $EW$ diagnostic diagrams for
\hd{36960}. All the observed \ioni{Si}{iii-iv} lines are included in the
plot. Uncertainties in the individual line abundances (propagated from the
errors in the measured $EW$s) are indicated as vertical lines\label{fig4}}
\end{figure}
%%%%%%%%%%%%%%%%%%%%%%%%%%%%%%%%%%%%%%%%%%%%%%%%%%%%%%%%%%%%%%%%%%%%%%%%%%%%

As mentioned, silicon abundance is obtained at the same time as
to the stellar parameters. Once a first estimation of the stellar parameters
was obtained through the $EW$(\ioni{Si}{iv}\,$\lambda$4116)/$EW$(\ioni{Si}{iii}\,$\lambda$4552)
and/or $EW$(\ioni{Si}{ii}\,$\lambda$6371)/$EW$(\ioni{Si}{iii}\,$\lambda$4552) line 
ratios, the curve-of-growth method was applied to a proper set of Si lines to 
derive the Si abundance, together with the final values of the stellar parameters 
and a microturbulence. 

Initially, we included all the measured lines in the analysis, but identified
some problematic lines. By {\em problematic lines} we mean lines giving abundances 
that are too high or too low compared with the mean value provide by a set of lines initially 
considered as reliable. These problematic lines were removed in the final analysis. 
To illustrate the procedure, we consider the case of \hd{36960}. The \abun{Si}-$EW$
diagram with whole set of measured Si lines (\ioni{Si}{iii} and \ioni{Si}{iv} lines, 
in this case) is shown in Fig. \ref{fig4}. As the initial set of reliable lines we consider
the \ioni{Si}{iii}\,$\lambda$$\lambda$4552, 4567, 4574 triplet and the \ioni{Si}{iv}\,$\lambda$4116 line.
The other \ioni{Si}{iv} lines provide similar abundances to the first set of reliable 
lines, except \ioni{Si}{iv}\,$\lambda$6701. The situation is less encouraging for the other
\ioni{Si}{iii} lines, since all of them lie below the mean abundance value from
reliable lines. We thus label these lines as problematic. The same 
procedure was followed for the selection of reliable \ioni{Si}{ii} lines (see Sect. 
\ref{parameters}).  

A similar behavior is found for the problematic lines in all the analyzed stars. Since 
the $EW$ of these lines are accurately measured and no lines from other 
elements are expected to be affecting them\footnote{This is not the case for 
\ioni{Si}{iv}\,$\lambda$$\lambda$4089, 4631, and \ioni{Si}{ii}\,$\lambda$$\lambda$6347, 3856; these lines may 
be blended with \ioni{O}{ii}\,$\lambda$4089, \ioni{N}{ii}\,$\lambda$4631, \ioni{Mg}{ii}\,$\lambda$6347, and
\ioni{O}{ii}$\lambda$3856, respectively, in some cases. The high resolution of the 
\fies\ spectra normally allows separating both line contributions; 
however, results with these lines are always treated with care.}, we argue 
that the discrepancy may be related to the definition of the model atom. 
For example, the \ioni{Si}{iii}\,$\lambda$$\lambda$4813, 4819, 4829 triplet is known to give different 
results than the \ioni{Si}{iii}\,$\lambda$$\lambda$4552, 4567, 4574 triplet because of the boundary 
problems of the \ioni{Si}{iii} model atom \citep{Bec90}. Normally, there is an 
explanation for the bad behavior of the problematic lines; therefore, the 
exclusion of lines from the analysis is not arbitrary (e.g. lines from the same
multiplets behave in the same way and are normally excluded all together).

It is important to notice that the analysis with the whole set of lines (including 
the problematic ones) could lead to an incorrect determination of the microturbulence, 
hence of the final value of Si abundance. 

Final results of the Si abundance analysis\footnote{Results of the line-by-line analysis 
are presented in Tables \ref{res1}-\ref{res13} (only available in the electronic
version).} are indicated in Table \ref{table3} (columns 9 and 10). Three sources of 
errors must be considered for estimating of the final uncertainty: (1) the dispersion 
in the line-by-line abundances, (2) the uncertainty associated with the microturbulence, 
(indicated in Table \ref{table3}, and Tables \ref{res1}-\ref{res13}, only available in 
the electronic version), and (3) the contribution of uncertainties in the stellar 
parameters. These can be added quadratically to obtain the total uncertainty.

For illustrative purposes in the last column of Table \ref{table3}, we indicate the effect of a change of 
$\pm$500 K in \Teff\ on the Si abundance if this was derived exclusively by using
\ioni{Si}{iii} lines\footnote{A similar test assuming a change of 0.1 dex in \grav\ 
indicates negligible effects when compared with the \Teff\ contribution.}. The minimum effect 
occurs for \Teff\,$\sim$\,27000\,K and increases
towards higher and lower temperatures (note the change in the sign of the uncertainties),
reaching values up to 0.12 dex.
This behavior is a consequence of the dependence of the $EW$ of the 
\ion{Si}{iii} lines with temperature. The maximum $EW$ is achieved around 27000\,K (so 
the abundance is quite insensitive to small changes in \Teff), and decreases
towards lower and higher temperatures (also increasing its sensitivity to \Teff\ variations).
When lines from two different ions are used, the final Si abundance 
can be constrained with better accuracy, since the $EW$ of lines from different ions behaves 
in a opposite way for a given \Teff. As a consequence in this case, the uncertainty associated 
with the stellar parameters is always negligible compared with the dispersion in the 
line-by-line abundances.

%=======================================================================
\subsubsection{Oxygen abundances}
%=======================================================================

% -----------------------------------------------------------------------
% --------------------           TABLE 4         ------------------------
% -----------------------------------------------------------------------
%
\begin{table*}[!t]
\begin{center}
\caption{Final results of the O abundance analysis. Detailed results (line-by-line) 
of the O abundance analysis are presented in Tables \ref{res1} to \ref{res13} in 
the electronic version of the paper.\label{table4}}
\begin{tabular}{ccccccccccc}
\noalign{\smallskip}
\tableline
\tableline
\noalign{\smallskip}
Target     &   SpT   & 
\Teff\ & \grav\ & \micro(O) & \abun{O} & $\Delta$\abun{O}($\sigma$) & $\Delta$\abun{O}(\micro) 
       & $\Delta$\abun{O}(\Teff$\pm$500) & \abun{O}(\Teff+500, \Teff-500) \\
\tableline
\noalign{\smallskip}
\hd{36512}  & B0\,V   & 33700 & 4.2 &  4.4$\pm$1.5 & {\bf 8.71} & 0.10 & 0.05 & $\pm$0.06 & (8.76, 8.65) \\
\hd{37020}  & B0.5\,V & 30500 & 4.2 &  6.4$\pm$1.6 & {\bf 8.70} & 0.10 & 0.07 & $\pm$0.05 & (8.74, 8.65)  \\
\hd{36960}  & B0.5\,V & 28900 & 3.9 &  5.9$\pm$0.8 & {\bf 8.71} & 0.10 & 0.04 & $\pm$0.03 & (8.74, 8.68)  \\
\hd{37042}  & B0.7\,V & 29700 & 4.2 &  4.9$\pm$1.1 & {\bf 8.75} & 0.08 & 0.06 & $\pm$0.02 & (8.78, 8.74)  \\
\hd{36591}  & B1\,V   & 27200 & 4.1 &  4.5$\pm$0.3 & {\bf 8.71} & 0.10 & 0.02 & $\mp$0.02 & (8.71, 8.74)  \\   
\hd{36959}  & B1\,V   & 25800 & 4.2 &  2.1$\pm$0.4 & {\bf 8.70} & 0.06 & 0.02 & $\mp$0.05 & (8.66, 8.76)  \\
\hd{37744}  & B1.5\,V & 23800 & 4.1 &  3.6$\pm$1.4 & {\bf 8.70} & 0.07 & 0.06 & $\mp$0.09 & (8.61, 8.79)  \\
\hd{35299}  & B1.5\,V & 23700 & 4.2 &  2.8$\pm$0.6 & {\bf 8.72} & 0.07 & 0.03 & $\mp$0.09 & (8.64, 8.82)  \\
\hd{36285}  & B2\,V   & 20600 & 4.0 &  5.5$\pm$1.5 & {\bf 8.80} & 0.10 & 0.06 & $\mp$0.13 & (8.67, 8.92)  \\
\hd{35039}  & B2\,V   & 19800 & 3.7 &  5.3$\pm$1.5 & {\bf 8.79} & 0.07 & 0.07 & $\mp$0.15 & (8.65, 8.94)  \\
\hd{36629}  & B2\,V   & 20000 & 4.1 &  6.0$\pm$1.7 & {\bf 8.76} & 0.10 & 0.06 & $\mp$0.14 & (8.62, 8.89)  \\
\hd{36430}  & B2\,V   & 18600 & 4.1 &  6.3$\pm$2.2 & {\bf 8.76} & 0.07 & 0.08 & $\mp$0.13 & (8.64, 8.90)  \\
\hd{35912}  & B2\,V   & 18500 & 4.0 &  6.3$\pm$2.2 & {\bf 8.79} & 0.09 & 0.08 & $\mp$0.13 & (8.67, 8.93)  \\
\noalign{\smallskip}
\tableline
\end{tabular}
\end{center}
\end{table*}
% -----------------------------------------------------------------------
% -----------------------------------------------------------------------

Up to 47 \ioni{O}{ii} lines were identified in the observed spectral range. However,
not all the lines were finally used for the O abundance determination. The selection 
of the final set of lines used for the analysis was based on a detailed 
analysis by multiplets (following similar criteria to the case of Si, but in
this case we studied the behavior of lines resulting from different multiplets). We
found a similar behavior of lines to the one described in \cite{Sim06}.

The derived O abundances are indicated in Table \ref{table4}, and the results of the
line-by-line analyses in Tables \ref{res1}-\ref{res13} (in the electronic version). 
For this element, only lines from one ionization state were available, so we could 
not test whether the corresponding ionization equilibrium (\ioni{O}{iii/ii} or \ioni{O}{ii/i}) is 
achieved for the considered stellar parameters. We adopt for the oxygen abundance
analysis the same stellar parameters as in the Si analysis. 

We found (as in many previous works) that the microturbulence derived from the
\ioni{O}{ii} lines (\micro(O)) differs from the Si analysis 
(\micro(Si)), and a somewhat larger microturbulence is derived. Some authors assume 
the \micro(Si) value (or a mean value of the microturbulences obtained for
the various elements analyzed) to perform the oxygen abundance analysis.
In our opinion, this can lead to significant systematic errors in the analysis.
Since this is an ad-hoc parameter that is still not well understood  
\citep[see, however,][]{Can09}, we adopt the microturbulence derived from
the oxygen analysis itself, for consistency. 

In fact, determination of the microturbulence value that will be adopted
in the final steps of the abundance determination is an important task. 
Unidentified problems in the \ioni{O}{ii} line modeling or bad measurements
of the corresponding $EW$s (due to blends, noise, or a bad placement
of the continuum) can enormously affect the \micro\ value that produces a zero 
slope in the \abun{O}-$EW$ diagrams. A detailed analysis by multiplets 
\citep[see][]{Sim05} can help identify problematic lines and to better decide on 
the final microturbulence to be adopted.

Table \ref{table4} also indicates the uncertainties associated with errors in
the line-to-line abundance dispersion, the microturbulence, and the stellar
parameters. In addition, last column shows the derived abundances if the
effective temperature is varied $\pm$500\,K. (Although the exact values are 
not shown here, the contribution of the \grav\ uncertainty to the oxygen 
abundance can be considered negligible in comparison with the \Teff\ 
contribution.) The derived oxygen 
abundance is very sensitive to changes in \Teff\ for the cooler and hotter
objects, and there is a change in the behavior of the oxygen abundance 
with \Teff\ around 27000\,K. This behavior is similar to the 
one illustrated in Table \ref{table3}.

%=======================================================================
\section{Comparison with previous works}\label{compare}
%=======================================================================

%%%%%%%%%%%%%%%%%%%%%%%%%%%%%%%%%%%%%%%%%%%%%%%%%%%%%%%%%%%%%%%%%%%%%%%%%%%%
\begin{figure*}[!t]
\centering
\includegraphics[angle=90,scale=0.60]{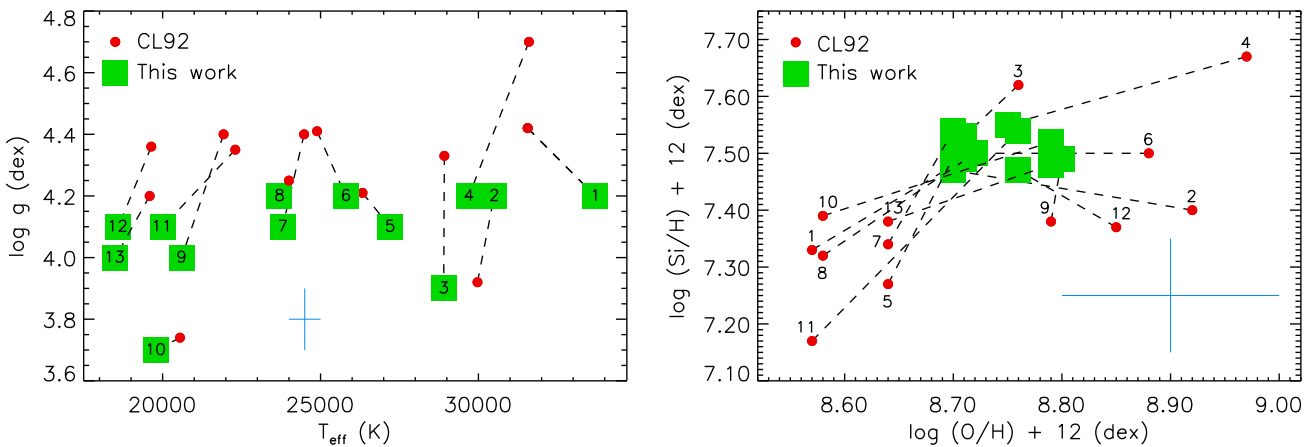}
\caption{Comparison of our derived stellar parameters, Si and O abundances,
with those obtained by \cite{Cun92, Cun94}. Results from both studies
for each star are connected by a dashed line. Numbers follow the order 
of stars presented in Tables \ref{table3} and \ref{table4}. 
The size of the uncertainties of the various quantities are shown by the
crosses. 
\label{fig5}}
\end{figure*}
%%%%%%%%%%%%%%%%%%%%%%%%%%%%%%%%%%%%%%%%%%%%%%%%%%%%%%%%%%%%%%%%%%%%%%%%%%%%

Two of the stars included in this analysis have already been analyzed in
\cite{Sim06}. These are \hd{37020} and \hd{37042}. 
We observed the two stars again to have spectra with the same
characteristics as the other stars in the sample. The stellar parameters 
presented in Table \ref{table2} correspond to the analysis of the new 
spectra. The new analysis resulted in slightly higher effective temperatures
(but within the errors) and gravities $\sim$\,0.1\,--\,0.2 dex larger. 
Several factors produced this difference in the derived stellar parameters.
First, we based our \Teff\,-\,\grav\ determination on the
HSi analysis, instead of using the H and \ioni{He}{i-ii} lines. The best 
\Teff\,-\,\grav\ pair reproducing the \ioni{Si}{iii-iv} ionization equilibrium 
and the wings of the H lines simultaneously is the one indicated
in Table \ref{table3}. A lower \grav\ requested an effective temperature
somewhat lower (which was not fitting the \ioni{He}{ii} lines).
In addition, the better quality of the new spectra allowed us to better 
constrain the gravity of the stars. The wings of the H lines 
are not very sensitive to changes of $\sim$\,0.1-0.2 dex in 
\grav\ in this range of stellar parameters. In \citeauthor{Sim06}, we based
our decision about the best solution on the faint 
\ioni{He}{ii}\,$\lambda$4541 line; however, it is better to rely on the HSi 
criterion, which is more sensitive in this range of stellar
parameters. In fact, we find that the new \Teff,\,-\,\grav\ pair
also fits nicely the \ioni{He}{i-ii} lines. We thus prefer this last solution.

We also obtain $\sim$0.05-0.10 dex higher oxygen abundances. Part of this
difference is caused by the change in stellar parameters; in addition, we 
measured sligthly larger $EW$s for the \ioni{O}{ii} lines in both stars. 
Note, however, that the old and new abundances agree when taking the error bars
into account.
The discrepancy for these two objects between \cite{Sim06} and 
these new studies serves as an example of the effect of intrinsic
uncertainties in determination of stellar abundances. Following a similar methodology
the same author can find slightly different results (but 
within the errors) when analyzing spectra from two different observing
campaigns\footnote{The different results may also be caused by actual changes
in the stellar spectra (e.g. due to binarity).}. \\

Recently, \cite{Prz08} analyzed a representative sample of unevolved
early B-type stars in nearby OB associations and the field using a similar
technique but a different stellar atmosphere code (line blanketed
{\sc atlas9} \lte\ model atmospheres, Kurucz 1993) and \nlte\ 
line-formation calculations (with updated versions of 
{\sc detail} and {\sc surface}, Giddings 1981; Butler \& Giddings 1985).
They have one star in common with our work, \hd{36591} (HR\,1861).
We obtain very similar results for the stellar parameters, as well as
the Si and O abundances. They used several \Teff\ spectroscopic 
indicators (apart from \ioni{Si}{ii-iv}), finding very good overall
agreement.

\cite{Gum98} included a sample of 5 stars from Ori\,OB1 in their study of the
abundance gradient of the Galaxy. We have three stars in common. \citeauthor{Gum98} 
used a self-consistent spectroscopic approach. We find similar effective temperatures 
for the two hotter objects (when \ioni{Si}{iv} lines are available), but $\sim$\,0.2\,dex
lower gravities. For these stars (\hd{36959} and \hd{36960}), the derived Si and 
O abundances agree (within the errors) with our values. The difference 
in gravity could explain their slightly lower O abundances. 

Interestingly, for the cooler object in common (\hd{35039}), we obtained lower \Teff\ 
and \grav. It is remarkable that they obtained very low O and Si abundances 
(8.20 and 7.08 dex, respectively) for this star. Probably, this result is related 
to the \ioni{Si}{ii} problem we described in Sect. \ref{parameters} (see also
lower panel in Fig. \ref{fig2}); they used 
line \ioni{Si}{ii}\,4130 to establish \Teff, so obtained too high a value 
(23500 K vs. 19900 K, the value we obtained). For this range of stellar parameters,
higher \Teff\ supposes lower O abundances, and a change of $\sim$3000 K in \Teff\ can
perfectly explain a 0.5 dex variation in the O abundance. \\

Figure \ref{fig5} presents a comparison our results with those obtained by 
\cite{Cun92, Cun94} concerning the stellar parameters, Si and O abundances.
There is a clear discrepancy for the majority of the objects, in terms not only  
of effective temperatures, but also of gravities. \cite{Cun92, Cun94} made use 
of the calibrations of Str\"omgen photometry coupled with the fits to the 
pressure-broadened line wings of H$\gamma$ \citep[from][LTE stellar atmopsheres]{Kur79} 
to derive \Teff\ and \grav. Following \cite{Gie92}, they adjusted the \Teff\ calibrations 
by \cite{Les86} and \cite{Bal84}
upward by 4.2\% and 5.2\% , respectively. The discrepancy in effective temperatures obtained
through commonly used photometric calibrations and spectroscopic line diagnostic 
has already been pointed out by several authors \citep[e.g.][]{Kil91, Nie08}. Our result 
once more brings out these discrepancies.

Note also the large discrepancy found for the gravity. The \citeauthor{Cun92} values are 
systematically higher ($\sim$0.3 dex). As shown by \cite{Nie07}, the use of 
\lte\ profiles for the gravity determination lead to overestimated \grav\ values, 
in particular for hotter stars. In Fig. \ref{fig3} we show a comparison of 
synthetic hydrogen lines from two \fastwind\ models with the observed profiles 
for \hd{36960} (labeled with \#3 in Fig.\,\ref{fig5}. One of the models considers
the stellar parameters derived from our analysis, the other one corresponds to the
\Teff-\grav\ pair provided by \cite{Cun92}. The difference in the wings of the lines
is clear.

\begin{figure*}[!t]
\centering
\includegraphics[angle=90,scale=.55]{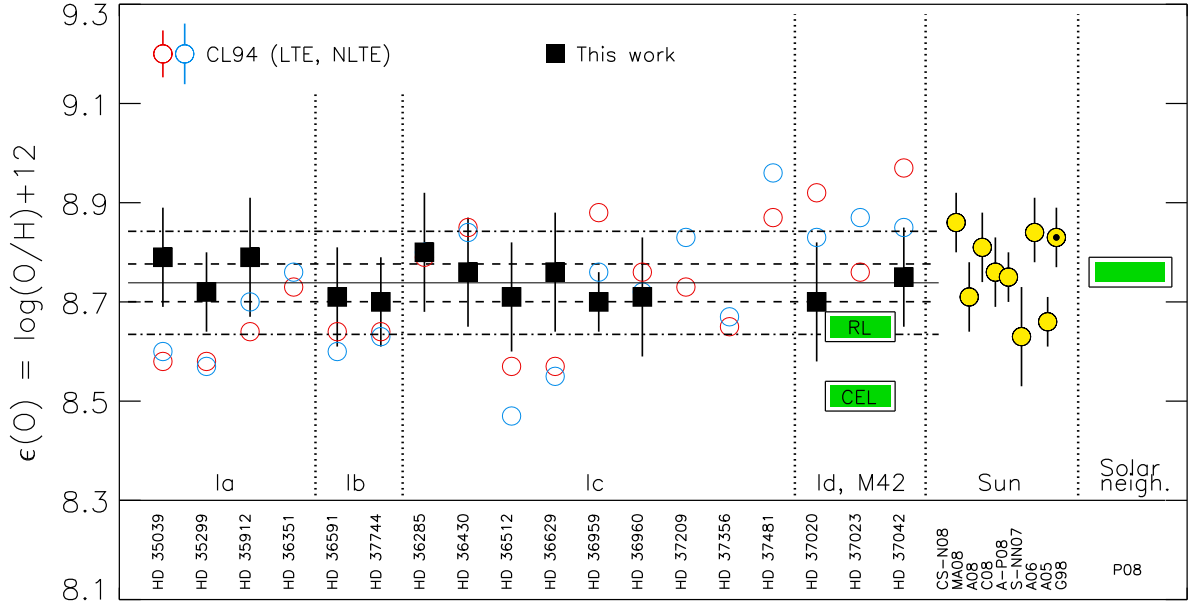}
\caption{Oxygen abundances derived for our sample of early B-type stars in the Orion
association, and comparison with the previous results from \cite{Cun94}. Vertical lines
separates stars from the various subgroups. Solid and dashed horizontal lines represent 
the mean value and the dispersion (1$\sigma$) of our results; dot-dashed horizontal
lines indicate the characteristic intrinsic uncertainty of the derived abundances
(accounting for uncertainties in stellar parameters, microturbulence, and the
line-by-line abundance dispersion). The oxygen
abundance derived by \cite{Est04} for M\,42, along with the values determined for the 
Solar abundance in the past 4 years (the epoch of the ``solar crisis"), are also 
presented for comparison. \label{fig6}}
\end{figure*}

We also find big discrepancies in the derived Si and O abundance for most 
of the stars, with no systematic trend (i.e. differences are found in 
both positive and negative directions).
There are several factors to take into account to explain this 
disagreement. One of them is the imprecise determination of the stellar
parameters from photometric indices. But maybe the most important one is the following.
Once the stellar parameters were calculated as described above, \citeauthor{Cun92} 
use those values to perform the abundance analysis, computing line-blanketed \lte\ model
atmospheres with {\sc atlas6} \citep{Kur79} for the \lte\ case, or using a grid
of $EW$s based on Gold's models \citep{Gol84} for the \nlte\ abundance calculations. The risk
in this strategy is that the stellar abundance analysis is decoupled of the
stellar parameter determination, and this could produce inconsistencies in the
analysis process. Briefly, the calculation of $EW$s of those lines used for 
the abundance determination is based on the stellar atmosphere structure
defined by a stellar atmosphere model computed for a given set of stellar 
parameters; on the other hand, the stellar parameters derived for a given star 
will depend on the characteristics of the stellar atmosphere model we have 
used\footnote{As an example, we want to mention the consequences that 
including the line blanketing and wind blanketing effects in the stellar 
atmosphere models of O and early B-type stars had on the SpT\,-\,\Teff\ 
calibrations of these stars \citep[see e.g.][]{Mar05, Rep04}.}. Therefore, it may be 
dangerous to do an 
abundance analysis with a given stellar atmosphere code using the stellar 
parameters obtained from a different code or a photometric calibration. This
argument is crucial in the case of stars with \Teff\,$\ge$\,30000 K, where photometric 
methods become completely unreliable discriminators of temperatures and gravities, 
because of the insensitivity  of the Rayleigh-Jeans tail of the spectral energy 
distribution on temperature.

A self-consistent spectroscopic approach allows minimizing this problem. In this 
case, the stellar parameters are determined by fitting certain spectroscopic 
diagnostics, and then the same models are used for the abundance analysis. This
way we are certain that the stellar atmosphere structure used for the computation
of the abundance diagnostics is coherent with the derived stellar parameters for 
the studied star. To a first order, when the whole analysis is performed with 
stellar atmosphere codes with different characteristics, the derived abundances 
should be quite similar (although the derived stellar parameters could be somewhat 
different). This approach can be strengthened if multiple independent 
spectroscopic indicators are considered \citep[i.e. \ioni{Si}{iii-iv}, \ioni{He}{i-ii}, 
\ioni{C}{ii-iv}; see e.g. this study or][]{Nie08}.

\begin{figure*}[!t]
\centering
\includegraphics[angle=90,scale=.55]{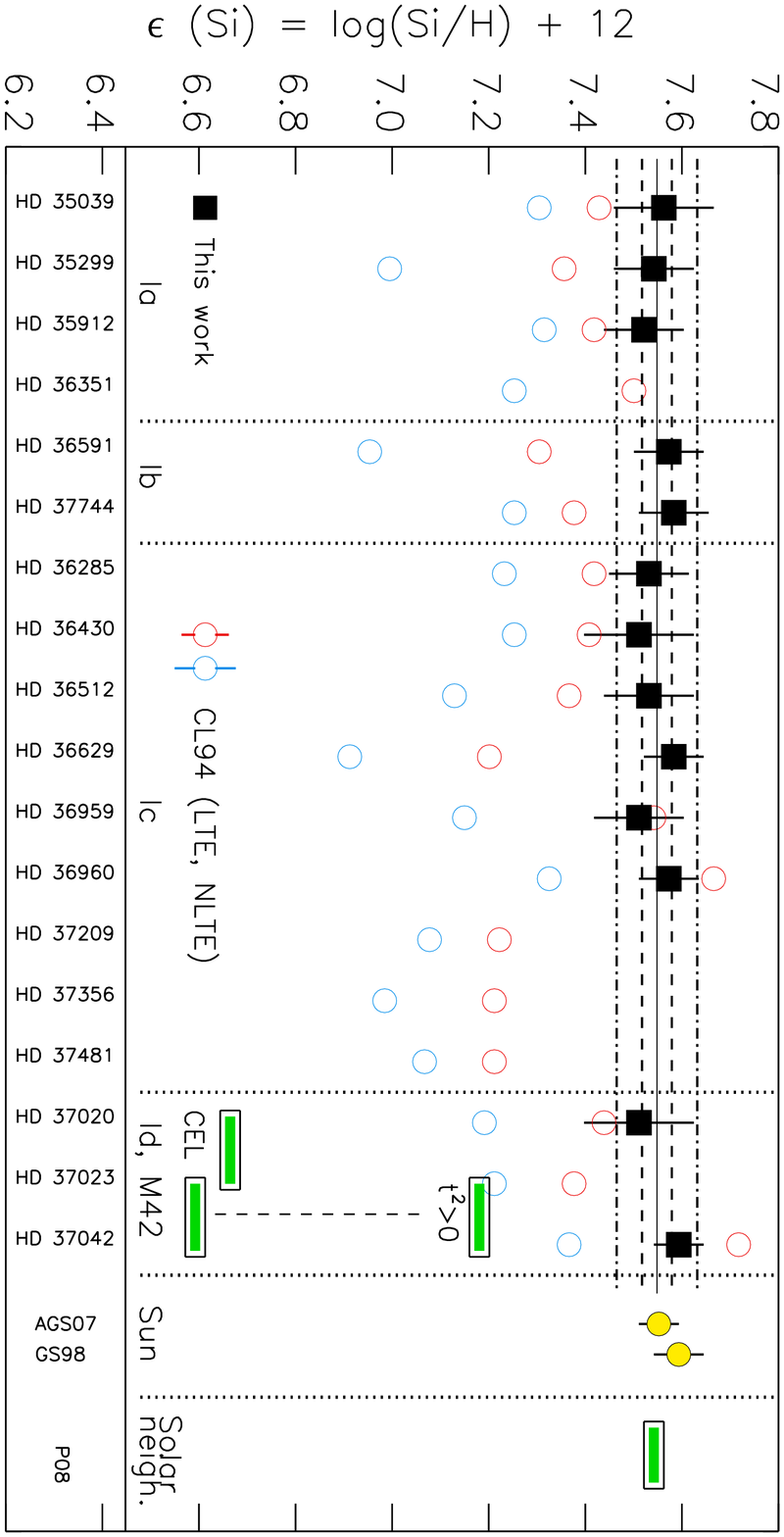}
\caption{As Fig. \ref{fig2} but for silicon. This time, the reference for
the nebular abundance is \cite{Rub93} and \cite{Gar95} (first and second column, see text).
Solar values from \cite{Asp05} and \cite{Gre98}\label{fig7}}
\end{figure*}

%=======================================================================
\section{Chemical composition of B-type stars in the Ori\,OB1 association}\label{discuss}
%=======================================================================

Figures \ref{fig6} and \ref{fig7} again show the derived oxygen and silicon
abundances in our sample of B-type stars in the Ori\,OB1 association, along
with the results by \cite{Cun94}. This time, the stars are ordered
following the subgroups suggested by \cite{Bla64}. As discussed in 
\cite{Cun94}, they found that the stars in the youngest group (Id) and some
of the stars in subgroup Ic were enriched in oxygen by about 40\% relative to 
the stars belonging to the older subgroups. They interpreted
this result as possible proof of enrichment of the new generation of
stars in the association with the products from supernovae ejecta 
from the older subgroups. In addition, they found features of this enrichment
in silicon (also expected from the triggered star formation scenario).
In contrast, we did not find any systematic difference between the O
and Si abundances in stars from the various associations. In fact, our results
indicate that the B-type stars in the Ori\,OB1 association are chemically
homogeneous (at least in terms of oxygen and silicon), having a dispersion in
abundances (0.04 and 0.03 dex, respectively) 
smaller than the intrinsic uncertainties of the derived abundances 
(0.10 and 0.08 dex, respectively).
The mean abundances are $\epsilon$(O)\,=\,8.73 dex and 
$\epsilon$(Si)\,=\,7.51 dex. These values agree with those 
obtained by \cite{Prz08} for their sample of six stars in the solar 
neighborhood. 

%=======================================================================
\subsection{Comparison with the Sun}
%=======================================================================

We also include in Fig. \ref{fig6} the resulting solar oxygen abundances 
appearing in the literature in the past 4 years (during the so-called solar 
oxygen crisis\footnote{\cite{Ayr06}} epoch, not yet finished). Various
authors have published values for the solar oxygen abundance based on
improved model atmospheres (either 1D and 3D), line formation codes, 
atomic and molecular data, and detailed treatment of blends. In the plot, 
we present results by \cite{Cen08}, \cite{Mel08}, \cite{Ayr08}, \cite{Caf08}, \cite{All08}, \cite{Soc07}, 
\cite{Ayr06}, \cite{Asp04}, and \cite{Gre98}. This last value (marked with an internal 
black dot) was considered as the standard solar O abundance until a few 
years ago. The derived solar values range from 8.63 dex \citep{Soc07} to 
8.86 dex \citep{Cen08}. The O abundances in our sample of stars in 
Ori\,OB1 lie in the middle of all these values. In view of the present-day
results, the only thing we can say is that oxygen abundances in the Sun
and B-type stars in the solar vicinity are the same within the uncertainties.
However, we consider it too premature to draw any firm conclusion or hypothesis
about the chemical evolution of the local interstellar medium during the
lifetime of the Sun.

Figure \ref{fig7} also shows a comparison of our derived Si abundances 
with the Solar value. We only present here the old abundance \citep{Gre98} 
and the new value proposed by \cite{Asp05}. In contrast to previous 
results, the mean Si abundance in B-type stars in Ori\,OB1 is very close to the
Solar value \citep[similar to other B-type stars in the solar vicinity][]{Prz08}.

%=======================================================================
\subsection{Comparison with the Orion nebula}
%=======================================================================

The most recent and detailed analysis of the optical spectrum of the Orion 
nebula was done by \cite{Est04}. The high-quality UVES@VLT spectrum they 
used allowed them to derive the oxygen gas phase abundance of the nebula 
by using collisionally excited lines (CELs) and recombination lines (RLs). 
The final value they propose is 8.65$\pm$0.03 dex. This value was 
calculated assuming the ionic abundances given by \ioni{O}{$^2+$} RLs and 
\ioni{O}{$^+$} CEL (plus a \tsq=0.022){\bf \footnote{The \tsq\ parameter 
was introduced by \cite{Pei67} to account for temperature 
fluctuations in ionized nebulae. The temperature fluctuation scenario has
been proposed to explain the CELs vs. RL abundance discrepancy 
\citep[see][and references therein]{Gar07}.}}. Note, however, the abundance 
given by the \ioni{O}{$^2+$} and \ioni{O}{$^+$} CELs (and \tsq=0) is 
8.51$\pm$0.03 dex. These values have been included in Fig. \ref{fig6} for 
comparison with the early B-type stellar abundances presented in this 
study. The mean value of the derived stellar oxygen abundances is 0.25 
and 0.11 dex higher than the nebular abundances given by the CELs and RLs, 
respectively. 

Although the mean stellar abundance seems to agree better with the one given
by the faint recombination lines, we have to consider that
the analysis of the nebular emission line spectrum can only provide
abundances for the ionized gas phase of the ISM. This means a lower
limit to the actual ISM abundance, since part of the oxygen can be 
depleted, forming part of the dust grains.

Silicon is one of the elements expected to be more depleted onto dust grains, 
along with Mg and Fe \citep{Dra03}. We can therefore compare our derived Si 
abundances with those obtained from the study of the emission line spectrum 
of the Orion nebula to try to find some clues about the amount of oxygen
depleted. One should note that the determination of Si nebular abundances is 
not straightforward and must be based on results from photoionization models 
and certain ionization correction factors (ICFs). The only determinations of 
Si abundance in the Orion nebula we found in the literature are those by 
\cite{Rub93} and \cite{Gar95}. Both studies used the same observations of 
the FUV \ioni{Si}{iii}]\,$\lambda\lambda$1883, 1892, and \ioni{C}{iii}]\,$\lambda\lambda$1907, 1909 
lines to estimate the Si abundance, and obtained 6.65 and 6.58 dex, respectively. 
They computed photoionization models to obtain Si/H \citep{Rub93} and the ICF 
needed to transform \ioni{Si}{$^{2+}$}/\ioni{C}{$^{2+}$} into Si/C, hence
the C/O and O/H ratios to derive Si/H \citep{Gar95}. In addition, \cite{Gar95}
discuss how the effect of assuming a \tsq=0.04 \citep{Pei93} in the
calculation of the C/O and O/H ratios could affect the derived Si abundance.
They obtained log\,(Si/H)+12\,=\,7.14 dex for this case. These three values 
of the Si abundance are included in Fig \ref{fig7} for comparison with the 
stellar abundances. The difference between stellar and gas phase Si abundances 
are $\sim$1 dex, or $\sim$0.3 dex, depending on the nebular abundances we trust 
more. 

Since a detailed comparison of stellar and nebular abundances in Ori\,OB1 
and the Orion nebula within a dust-depletion scenario requires more extended 
study, which is beyond the scope of this paper, we decide to present a more 
detailed discussion in a separate paper (in preparation).

%=======================================================================
\section{Summary and conclusion}\label{conclude}
%=======================================================================

In this work, we applied a self-consistent spectroscopic approach to 
determining of the stellar parameters, the Si and O abundances 
of a sample of early-B type stars from the various subgroups of the 
Ori\,OB1 association. We made use of a high-quality spectroscopic
data\,set, obtained with \fies, and the modern \nlte,
line-blanketed, spherically extended stellar atmosphere code \fastwind.

We developed several IDL programs to automatically identify and measure
the $EW$ of metal lines in high-resolution spectra. We also
constructed a grid of \fastwind\ HHeSiO models optimized for the analysis
of early B-type, main sequence stars.

The availability of a large number of \ioni{Si}{ii-iii-iv} lines in the
\fies\ spectra allowed
us to obtain the stellar parameters with high accuracy and detect some
problems related to some Si lines commonly used for the stellar 
parameter and Si abundance determination. Once these problems were 
accounted for, a high degree of homogeneity was found for Si abundances
in the analyzed sample of stars.

The oxygen abundance analysis also result in a small dispersion of 
abundances, in contrast to previous determinations.
The mean oxygen and silicon abundances agree with those resulting
from a similar analysis of a representative sample of unevolved 
early B-type stars in the solar vicinity \citep{Prz08}. Both results
indicate that abundances derived from these stellar objects are
more homogeneous and metal-rich than previously though.

We also compared the O and Si stellar abundance in Ori\,OB1 with those
obtained for the Sun during the epoch of the ``solar  crisis". The O 
abundances in our sample of stars in Ori\,OB1 lie in the middle of 
all these values. In view of the present-day results, the only thing we 
can say is that oxygen abundances in the Sun and B-type stars in the 
solar vicinity are the same within the uncertainties. However, we 
consider it too premature to draw any firm conclusion or hypothesis
about the chemical evolution of the local interstellar medium during the
lifetime of the Sun. Silicon abundances are also very similar (contrary
to what was previously found from the study of B-type stars).

Finally, we compared the stellar abundances with those derived from the
study of the emission line spectrum of the Orion nebula. In a forthcoming
paper we will present a more detailed discussion accounting for the possible
depletion of O, Si, Mg, and Fe into dust.

This work points out one more time the importance of self-consistent 
spectroscopic abundance analyses for determining of the chemical
composition of the photospheres of OB-type stars. Photometric \Teff\ 
diagnostics must be treated with caution in the context of abundance
analyses of these type of objects. It is also dangerous to combine the
stellar parameters determined with any given code and the abundance
analysis performed with a different code. Systematic errors inherent 
in those techniques, or possible biases between the various codes, can 
lead to incorrect abundances.

%=======================================================================
\begin{acknowledgements}
%=======================================================================

Financial support by the Spanish Ministerio de Ciencia e Innovaci\'on under
the project AYA2008-06166-C03-01. This work has also been partially
funded by the Spanish MICINN under the Consolider-Ingenio 2010 Program 
grant CSD2006-00070: First Science with the GTC  (http://www.iac.es/consolider-ingenio-gtc).
I am very grateful to A. Herrero (Spain), G. Stasinska (France), and 
D. Schaerer (Switzerland) for their support and hospitality during the 
development of this work. I also acknowledge A. Herrero, M. Urbaneja, D. 
Lennon, F. Najarro, C. Trundle, and N. Castro for fruitful discussions. 
J. Puls, N. Przybilla, F. Nieva, and M. Urbaneja for the careful 
reading of the first version of this paper and all their comments. Finally, 
I thank J. Puls for allowing me to use the stellar atmosphere code \fastwind.

\end{acknowledgements}

%=======================================================================

\Online

% -----------------------------------------------------------------------
% -----------------------------------------------------------------------
%
\begin{table}[!t]
\begin{center}
\caption{List of log\,$gf$ values for the \ioni{Si}{ii-iv} lines considered 
in this study (from Atomic Line List v2.05). For the \ioni{O}{ii} lines we 
refer to \cite{Sim06}\label{tab-on1}}
\begin{tabular}{ccccc}
\noalign{\smallskip}
\tableline
\tableline
\noalign{\smallskip}
Ion & $\lambda$ \AA & log\,$gf$ & Configuration & Term \\
\tableline
\noalign{\smallskip}
\ioni{Si}{ii} & 3856.02 & -0.4569 & 3s 3p$^2$-3s$^2$ (1S) 4p & 2D-2Po \\
\ioni{Si}{ii} & 3862.60 & -0.7130 & 3s 3p$^2$-3s$^2$ (1S) 4p & 2D-2Po \\ 
\ioni{Si}{ii} & 4128.10 &  0.3646 & 3s$^2$ (1S) 3d-3s$^2$ (1S) 4f & 2D-2Fo \\ 
\ioni{Si}{ii} & 4130.90 &  0.5192 & 3s$^2$ (1S) 3d-3s$^2$ (1S) 4f & 2D-2Fo \\
\ioni{Si}{ii} & 5041.02 &  0.2543 & 3s$^2$ (1S) 4p-3s$^2$ (1S) 4d & 2Po-2D \\
\ioni{Si}{ii} & 5055.98 &  0.5083 & 3s$^2$ (1S) 4p-3s$^2$ (1S) 4d & 2Po-2D \\
\ioni{Si}{ii} & 5056.32 & -0.4460 & 3s$^2$ (1S) 4p-3s$^2$ (1S) 4d & 2Po-2D \\
\ioni{Si}{ii} & 6347.11 &  0.1819 & 3s$^2$ (1S) 4s-3s$^2$ (1S) 4p & 2S-2Po \\
\ioni{Si}{ii} & 6371.37 & -0.1208 & 3s$^2$ (1S) 4s-3s$^2$ (1S) 4p & 2S-2Po \\
\noalign{\smallskip}
\ioni{Si}{iii} & 4552.62 &  0.2828 & 3s 4s-3s 4p & 3S-3Po \\
\ioni{Si}{iii} & 4567.84 &  0.0595 & 3s 4s-3s 4p & 3S-3Po \\
\ioni{Si}{iii} & 4574.76 & -0.4183 & 3s 4s-3s 4p & 3S-3Po \\
\ioni{Si}{iii} & 5739.73 & -0.1032 & 3s 4s-3s 4p & 1S-1Po \\
\noalign{\smallskip}
\ioni{Si}{iv} & 4088.86 &  0.1984 & 2p$^6$ 4s-2p$^6$ 4p & 2S-2Po \\
\ioni{Si}{iv} & 4116.10 & -0.1055 & 2p$^6$ 4s-2p$^6$ 4p & 2S-2Po \\
\ioni{Si}{iv} & 4212.40 &  0.3840 & 2p$^6$ 5d-2p$^6$ 6f & 2D-2Fo \\
\ioni{Si}{iv} & 4212.41 &  0.5601 & 2p$^6$ 5d-2p$^6$ 6f & 2D-2Fo \\
\ioni{Si}{iv} & 4631.24 &  1.2158 & 2p$^6$ 5f-2p$^6$ 6g & 2Fo-2G \\
\noalign{\smallskip}
\tableline
\end{tabular}
\end{center}
\end{table}
% -----------------------------------------------------------------------
% -----------------------------------------------------------------------

.\\ \newpage 
.\\ \newpage 

% -----------------------------------------------------------------------
% --------------------           HD36512         ------------------------
% -----------------------------------------------------------------------
%
\begin{table}[!h]
{\scriptsize
\begin{center}
%----------------------
\begin{minipage}[c]{0.45\textwidth}
\caption{\scriptsize Results from the abundance analysis of 
%----------------------
HD\,36512 (B0\,V) 
\label{res1}}
\begin{tabular}{ccccc}
\noalign{\smallskip}
\tableline\tableline
\noalign{\smallskip}
%----------------------
HD\,36512 & 
\multicolumn{3}{c}{\Teff\,=\,33700 K, \grav\,=\,4.2 dex} & \micro(Si)=4.3  \\
%----------------------
\tableline
\noalign{\smallskip}
Line & $EW$ & $\Delta$$EW$ & $\epsilon_{\rm Si}$ & $\Delta\epsilon_{\rm Si}$ \\
 & (m\AA) & (m\AA) & (dex) & (dex) \\
\tableline
\noalign{\smallskip}
%----------------------
SiIII4552 & 117 &  4 & 7.51 & 0.07 \\
SiIII4567 &  99 &  5 & 7.58 & 0.08 \\
SiIII4574 &  50 &  4 & 7.53 & 0.09 \\
 SiIV4089 & 189 &  5 & 7.42 & 0.06 \\
 SiIV4116 & 166 &  5 & 7.54 & 0.07 \\
 SiIV4212 &  53 &  3 & 7.46 & 0.06 \\
 SiIV4631 &  75 &  4 & 7.40 & 0.06 \\
%----------------------
\tableline
\noalign{\smallskip}
\multicolumn{3}{c}{} & 
%----------------------
 {\bf $\epsilon_{\rm Si}$\,=\, 7.49} & 
 \multicolumn{1}{r}{$\Delta\epsilon_{\rm Si}$($\sigma$)\,=\,0.07} \\
 &  \multicolumn{2}{r}{$\Delta$\micro(Si)\,=\,0.7}  & $\Rightarrow$ &
 \multicolumn{1}{r}{$\Delta\epsilon_{\rm Si}$(\micro)\,=\,0.05} \\
%----------------------
\end{tabular}
\end{minipage}  
\begin{minipage}[c]{0.45\textwidth}
\begin{tabular}{ccccc}
\noalign{\smallskip}
\tableline\tableline
\noalign{\smallskip}
%----------------------
HD\,36512 & 
\multicolumn{3}{c}{\Teff\,=\,33700 K, \grav\,=\,4.2 dex} & \micro(O)=4.4 \\
%----------------------
\tableline
\noalign{\smallskip}
Line & $EW$ & $\Delta$$EW$ & $\epsilon_{\rm O}$ & $\Delta\epsilon_{\rm O}$ \\
 & (m\AA) & (m\AA) & (dex) & (dex) \\
\tableline
\noalign{\smallskip}
%----------------------
  OII3945 &  36 &  5 & 8.81 & 0.13 \\
  OII3954 &  63 &  5 & 8.86 & 0.09 \\
  OII3982 &  29 &  5 & 8.63 & 0.14 \\
  OII4317 &  78 & 11 & 8.74 & 0.16 \\
  OII4319 &  75 &  5 & 8.77 & 0.08 \\
  OII4366 &  70 &  8 & 8.70 & 0.14 \\
  OII4414 &  89 &  6 & 8.55 & 0.09 \\
  OII4416 &  76 &  9 & 8.66 & 0.15 \\
  OII4452 &  25 & 10 & 8.59 & 0.33 \\
  OII4641 &  95 & 12 & 8.69 & 0.20 \\
  OII4661 &  82 &  4 & 8.88 & 0.07 \\
  OII4673 &  23 &  3 & 8.83 & 0.12 \\
  OII4676 &  61 &  5 & 8.67 & 0.10 \\
  OII4696 &  13 &  6 & 8.73 & 0.37 \\
  OII6721 &  30 &  8 & 8.71 & 0.20 \\
  OII4072 &  84 &  8 & 8.62 & 0.15 \\
  OII4076 & 101 &  6 & 8.64 & 0.11 \\
  OII4078 &  33 &  6 & 8.73 & 0.18 \\
  OII4086 &  48 &  8 & 8.88 & 0.20 \\
  OII4089 &  77 &  6 & 8.63 & 0.10 \\
  OII4891 &  19 &  5 & 8.65 & 0.20 \\
  OII4906 &  34 &  5 & 8.67 & 0.13 \\
  OII4943 &  46 &  3 & 8.59 & 0.06 \\
%----------------------
\tableline
\noalign{\smallskip}
 \multicolumn{3}{c}{} & 
%----------------------
 {\bf $\epsilon_{\rm O}$\,=\, 8.71} & 
 \multicolumn{1}{r}{$\Delta\epsilon_{\rm O}$($\sigma$)\,=\,0.10} \\
 &  \multicolumn{2}{r}{$\Delta$\micro(O)\,=\,1.5}  & $\Rightarrow$ &
 \multicolumn{1}{r}{$\Delta\epsilon_{\rm O}$(\micro)\,=\,0.05} \\
% \multicolumn{3}{r}{$\Delta$\Teff\,=\,500, $\Delta$\grav\,=\,0.1} & $\Rightarrow$ & 
% \multicolumn{1}{r}{$\Delta\epsilon_{\rm O}$(SP)\,=\,0.06} \\
%----------------------
\end{tabular}
\end{minipage}  
%% Any table notes must follow the \end{tabular} command.
%\tablecomments{\footnotesize }
\end{center}
}\end{table}
%
% -----------------------------------------------------------------------
% -----------------------------------------------------------------------

% -----------------------------------------------------------------------
% --------------------           HD37020         ------------------------
% -----------------------------------------------------------------------
%
\begin{table}[!h]
{\scriptsize
\begin{center}
%----------------------
\begin{minipage}[c]{0.45\textwidth}
\caption{\scriptsize Results from the abundance analysis of 
%----------------------
HD\,37020 (B0.5\,V) 
\label{res2}}
%----------------------
\begin{tabular}{ccccc}
\noalign{\smallskip}
\tableline\tableline
\noalign{\smallskip}
%----------------------
HD\,37020 & 
\multicolumn{3}{c}{\Teff\,=\,30500 K, \grav\,=\,4.2 dex} & \micro(Si)=0.5  \\
%----------------------
\tableline
\noalign{\smallskip}
Line & $EW$ & $\Delta$$EW$ & $\epsilon_{\rm Si}$ & $\Delta\epsilon_{\rm Si}$ \\
 & (m\AA) & (m\AA) & (dex) & (dex) \\
\tableline
\noalign{\smallskip}
%----------------------
SiIII4552 & 130 & 13 & 7.56 & 0.27 \\
SiIII4567 & 110 &  7 & 7.54 & 0.14 \\
SiIII4574 &  74 &  9 & 7.57 & 0.22 \\
SiIII5739 &  73 &  9 & 7.38 & 0.19 \\
 SiIV4089 & 104 & 21 & 7.29 & 0.42 \\
 SiIV4116 &  90 &  8 & 7.41 & 0.17 \\
 SiIV4212 &  25 &  5 & 7.43 & 0.21 \\
 SiIV4631 &  47 &  7 & 7.55 & 0.17 \\
%----------------------
\tableline
\noalign{\smallskip}
\multicolumn{3}{c}{} & 
%----------------------
 {\bf $\epsilon_{\rm Si}$\,=\, 7.47} & 
 \multicolumn{1}{r}{$\Delta\epsilon_{\rm Si}$($\sigma$)\,=\,0.10} \\
 &  \multicolumn{2}{r}{$\Delta$\micro(Si)\,=\,0.5}  & $\Rightarrow$ &
 \multicolumn{1}{r}{$\Delta\epsilon_{\rm Si}$(\micro)\,=\,0.04} \\
%----------------------
\end{tabular}
\end{minipage}  
\begin{minipage}[c]{0.45\textwidth}
\begin{tabular}{ccccc}
\noalign{\smallskip}
\tableline\tableline
\noalign{\smallskip}
%----------------------
HD\,37020 &
\multicolumn{3}{c}{\Teff\,=\,30500 K, \grav\,=\,4.2 dex} & \micro(O)=6.4 \\
%----------------------
\tableline
\noalign{\smallskip}
%----------------------
  OII3945 &  56 & 10 & 8.75 & 0.19 \\
  OII3954 &  90 & 10 & 8.77 & 0.15 \\
  OII4317 &  94 & 10 & 8.56 & 0.13 \\
  OII4319 & 106 & 13 & 8.73 & 0.18 \\
  OII4366 & 104 & 13 & 8.71 & 0.19 \\
  OII4414 & 130 & 10 & 8.50 & 0.12 \\
  OII4416 & 119 & 10 & 8.68 & 0.13 \\
  OII4452 &  62 & 12 & 8.83 & 0.21 \\
  OII4638 & 101 & 10 & 8.83 & 0.15 \\
  OII4641 & 141 & 10 & 8.73 & 0.14 \\
  OII4661 & 105 & 10 & 8.71 & 0.15 \\
  OII4676 &  94 & 12 & 8.67 & 0.19 \\
  OII6721 &  61 & 12 & 8.77 & 0.19 \\
  OII4076 & 144 & 10 & 8.73 & 0.15 \\
  OII4891 &  40 &  8 & 8.84 & 0.19 \\
  OII4906 &  48 &  9 & 8.63 & 0.18 \\
  OII4941 &  48 &  6 & 8.57 & 0.12 \\
  OII4943 &  65 &  8 & 8.53 & 0.13 \\
%----------------------
\tableline
\noalign{\smallskip}
 \multicolumn{3}{c}{} & 
%----------------------
 {\bf $\epsilon_{\rm O}$\,=\, 8.70} & 
 \multicolumn{1}{r}{$\Delta\epsilon_{\rm O}$($\sigma$)\,=\,0.10} \\
 &  \multicolumn{2}{r}{$\Delta$\micro(O)\,=\,1.6}  & $\Rightarrow$ &
 \multicolumn{1}{r}{$\Delta\epsilon_{\rm O}$(\micro)\,=\,0.07} \\
% \multicolumn{3}{r}{$\Delta$\Teff\,=\,500, $\Delta$\grav\,=\,0.1} & $\Rightarrow$ & 
% \multicolumn{1}{r}{$\Delta\epsilon_{\rm O}$(SP)\,=\,0.05} \\
%----------------------
\end{tabular}
\end{minipage}  
%% Any table notes must follow the \end{tabular} command.
%\tablecomments{\footnotesize }
\end{center}
}\end{table}
%
% -----------------------------------------------------------------------
% -----------------------------------------------------------------------

% -----------------------------------------------------------------------
% --------------------           HD36960         ------------------------
% -----------------------------------------------------------------------
%
\begin{table}[!h]
{\scriptsize
\begin{center}
%----------------------
\begin{minipage}[c]{0.45\textwidth}
\caption{\scriptsize Results from the abundance analysis of 
%----------------------
HD\,36960 (B0.5\,V) 
\label{res3}}
%----------------------
\begin{tabular}{ccccc}
\noalign{\smallskip}
\tableline\tableline
\noalign{\smallskip}
%----------------------
HD\,36960 & 
\multicolumn{3}{c}{\Teff\,=\,28900 K, \grav\,=\,3.9 dex} & \micro(Si)=5.4  \\
%----------------------
\tableline
\noalign{\smallskip}
Line & $EW$ & $\Delta$$EW$ & $\epsilon_{\rm Si}$ & $\Delta\epsilon_{\rm Si}$ \\
 & (m\AA) & (m\AA) & (dex) & (dex) \\
\tableline
\noalign{\smallskip}
%----------------------
SiIII4552 & 211 &  5 & 7.53 & 0.06 \\
SiIII4567 & 179 &  5 & 7.51 & 0.06 \\
SiIII4574 & 116 &  5 & 7.52 & 0.07 \\
 SiIV4089 & 160 & 11 & 7.55 & 0.15 \\
 SiIV4116 & 127 &  9 & 7.52 & 0.14 \\
 SiIV4212 &  28 &  4 & 7.50 & 0.14 \\
 SiIV4631 &  37 &  3 & 7.54 & 0.08 \\
%----------------------
\tableline
\noalign{\smallskip}
\multicolumn{3}{c}{} & 
%----------------------
 {\bf $\epsilon_{\rm Si}$\,=\, 7.53} & 
 \multicolumn{1}{r}{$\Delta\epsilon_{\rm Si}$($\sigma$)\,=\,0.02} \\
 &  \multicolumn{2}{r}{$\Delta$\micro(Si)\,=\,0.6}  & $\Rightarrow$ &
 \multicolumn{1}{r}{$\Delta\epsilon_{\rm Si}$(\micro)\,=\,0.06} \\
%----------------------
\end{tabular}
\end{minipage}  
\begin{minipage}[c]{0.45\textwidth}
\begin{tabular}{ccccc}
\noalign{\smallskip}
\tableline\tableline
\noalign{\smallskip}
%----------------------
HD\,36960 & 
\multicolumn{3}{c}{\Teff\,=\,28900 K, \grav\,=\,3.9 dex} & \micro(O)=5.9 \\
%----------------------
\tableline
\noalign{\smallskip}
%----------------------
  OII3945 &  68 &  6 & 8.78 & 0.11 \\
  OII3954 &  96 &  7 & 8.71 & 0.10 \\
  OII3982 &  61 &  6 & 8.63 & 0.11 \\
  OII4317 & 125 &  7 & 8.71 & 0.08 \\
  OII4319 & 125 &  6 & 8.78 & 0.08 \\
  OII4366 & 113 &  5 & 8.66 & 0.07 \\
  OII4414 & 165 &  5 & 8.66 & 0.06 \\
  OII4416 & 141 &  5 & 8.75 & 0.06 \\
  OII4452 &  50 &  5 & 8.54 & 0.10 \\
  OII4661 & 127 &  5 & 8.87 & 0.08 \\
  OII4673 &  46 &  5 & 8.85 & 0.12 \\
  OII4676 & 101 &  5 & 8.66 & 0.08 \\
  OII4696 &  30 &  3 & 8.78 & 0.09 \\
  OII6641 &  47 &  6 & 8.81 & 0.12 \\
  OII6721 &  76 &  9 & 8.77 & 0.12 \\
  OII4078 &  60 &  4 & 8.81 & 0.09 \\
  OII4891 &  34 & 10 & 8.64 & 0.27 \\
  OII4906 &  61 &  6 & 8.71 & 0.11 \\
  OII4941 &  54 & 10 & 8.54 & 0.19 \\
  OII4943 &  75 &  4 & 8.54 & 0.06 \\
  OII4956 &  23 &  4 & 8.72 & 0.14 \\
%----------------------
\tableline
\noalign{\smallskip}
 \multicolumn{3}{c}{} & 
%----------------------
 {\bf $\epsilon_{\rm O}$\,=\, 8.71} & 
 \multicolumn{1}{r}{$\Delta\epsilon_{\rm O}$($\sigma$)\,=\,0.10} \\
 &  \multicolumn{2}{r}{$\Delta$\micro(O)\,=\,0.8}  & $\Rightarrow$ &
 \multicolumn{1}{r}{$\Delta\epsilon_{\rm O}$(\micro)\,=\,0.04} \\
% \multicolumn{3}{r}{$\Delta$\Teff\,=\,500, $\Delta$\grav\,=\,0.1} & $\Rightarrow$ & 
% \multicolumn{1}{r}{$\Delta\epsilon_{\rm O}$(SP)\,=\,0.03} \\
%----------------------
\end{tabular}
\end{minipage}  
%% Any table notes must follow the \end{tabular} command.
%\tablecomments{\footnotesize }
\end{center}
}\end{table}
%
% -----------------------------------------------------------------------
% -----------------------------------------------------------------------

% -----------------------------------------------------------------------
% --------------------           HD37042         ------------------------
% -----------------------------------------------------------------------
%
\begin{table}[!h]
{\scriptsize
\begin{center}
%----------------------
\begin{minipage}[c]{0.45\textwidth}
\caption{\scriptsize Results from the abundance analysis of 
%----------------------
HD\,37042 (B0.7\,V) 
\label{res4}}
%----------------------
\begin{tabular}{ccccc}
\noalign{\smallskip}
\tableline\tableline
\noalign{\smallskip}
%----------------------
HD\,37042 & 
\multicolumn{3}{c}{\Teff\,=\,29700 K, \grav\,=\,4.2 dex} & \micro(Si)=1.4  \\
%----------------------
\tableline
\noalign{\smallskip}
Line & $EW$ & $\Delta$$EW$ & $\epsilon_{\rm Si}$ & $\Delta\epsilon_{\rm Si}$ \\
 & (m\AA) & (m\AA) & (dex) & (dex) \\
\tableline
\noalign{\smallskip}
%----------------------
SiIII4552 & 149 &  2 & 7.54 & 0.04 \\
SiIII4567 & 133 &  6 & 7.59 & 0.11 \\
SiIII4574 &  88 &  4 & 7.56 & 0.08 \\
SiIII5739 &  99 &  7 & 7.50 & 0.13 \\
 SiIV4116 &  95 & 10 & 7.53 & 0.21 \\
 SiIV4212 &  24 &  5 & 7.54 & 0.22 \\
 SiIV4631 &  35 & 10 & 7.55 & 0.30 \\
%----------------------
\tableline
\noalign{\smallskip}
\multicolumn{3}{c}{} & 
%----------------------
 {\bf $\epsilon_{\rm Si}$\,=\, 7.55} & 
 \multicolumn{1}{r}{$\Delta\epsilon_{\rm Si}$($\sigma$)\,=\,0.03} \\
 &  \multicolumn{2}{r}{$\Delta$\micro(Si)\,=\,0.3}  & $\Rightarrow$ &
 \multicolumn{1}{r}{$\Delta\epsilon_{\rm Si}$(\micro)\,=\,0.04} \\
%----------------------
\end{tabular}
\end{minipage}  
\begin{minipage}[c]{0.45\textwidth}
\begin{tabular}{ccccc}
\noalign{\smallskip}
\tableline\tableline
\noalign{\smallskip}
%----------------------
HD\,37042 & 
\multicolumn{3}{c}{\Teff\,=\,29700 K, \grav\,=\,4.2 dex} & \micro(O)=4.9 \\
%----------------------
\tableline
\noalign{\smallskip}
%----------------------
  OII3945 &  68 &  6 & 8.87 & 0.11 \\
  OII3954 &  93 &  7 & 8.79 & 0.12 \\
  OII4317 & 116 & 10 & 8.75 & 0.13 \\
  OII4319 & 116 & 11 & 8.84 & 0.16 \\
  OII4366 & 108 &  9 & 8.76 & 0.14 \\
  OII4414 & 143 & 10 & 8.64 & 0.13 \\
  OII4416 & 127 &  7 & 8.78 & 0.10 \\
  OII4452 &  64 &  6 & 8.82 & 0.12 \\
  OII4641 & 142 &  7 & 8.81 & 0.11 \\
  OII4661 & 108 &  5 & 8.79 & 0.09 \\
  OII4676 &  95 &  6 & 8.71 & 0.11 \\
  OII4696 &  30 &  5 & 8.82 & 0.16 \\
  OII4072 & 110 & 10 & 8.62 & 0.18 \\
  OII4076 & 128 & 10 & 8.63 & 0.17 \\
  OII4078 &  57 &  8 & 8.84 & 0.19 \\
  OII4891 &  39 &  5 & 8.80 & 0.13 \\
  OII4906 &  57 &  5 & 8.75 & 0.10 \\
  OII4941 &  53 &  4 & 8.63 & 0.08 \\
  OII4943 &  73 &  4 & 8.63 & 0.07 \\
  OII4956 &  23 &  8 & 8.79 & 0.29 \\
%----------------------
\tableline
\noalign{\smallskip}
 \multicolumn{3}{c}{} & 
%----------------------
 {\bf $\epsilon_{\rm O}$\,=\, 8.75} & 
 \multicolumn{1}{r}{$\Delta\epsilon_{\rm O}$($\sigma$)\,=\,0.08} \\
 &  \multicolumn{2}{r}{$\Delta$\micro(O)\,=\,1.1}  & $\Rightarrow$ &
 \multicolumn{1}{r}{$\Delta\epsilon_{\rm O}$(\micro)\,=\,0.06} \\
% \multicolumn{3}{r}{$\Delta$\Teff\,=\,500, $\Delta$\grav\,=\,0.1} & $\Rightarrow$ & 
% \multicolumn{1}{r}{$\Delta\epsilon_{\rm O}$(SP)\,=\,0.02} \\
%----------------------
\end{tabular}
\end{minipage}  
%% Any table notes must follow the \end{tabular} command.
%\tablecomments{\footnotesize }
\end{center}
}\end{table}
%
% -----------------------------------------------------------------------
% -----------------------------------------------------------------------

% -----------------------------------------------------------------------
% --------------------           HD36591         ------------------------
% -----------------------------------------------------------------------
%
\begin{table}[!h]
{\scriptsize
\begin{center}
%----------------------
\begin{minipage}[c]{0.45\textwidth}
\caption{\scriptsize Results from the abundance analysis of 
%----------------------
HD\,36591 (B1\,V) 
\label{res5}}
%----------------------
\begin{tabular}{ccccc}
\noalign{\smallskip}
\tableline\tableline
\noalign{\smallskip}
%----------------------
HD\,36591 & 
\multicolumn{3}{c}{\Teff\,=\,27200 K, \grav\,=\,4.1 dex} & \micro(Si)=1.3  \\
%----------------------
\tableline
\noalign{\smallskip}
Line & $EW$ & $\Delta$$EW$ & $\epsilon_{\rm Si}$ & $\Delta\epsilon_{\rm Si}$ \\
 & (m\AA) & (m\AA) & (dex) & (dex) \\
\tableline
\noalign{\smallskip}
%----------------------
SiIII4552 & 158 &  2 & 7.52 & 0.03 \\
SiIII4567 & 145 &  2 & 7.62 & 0.04 \\
SiIII4574 & 104 &  2 & 7.65 & 0.04 \\
SiIII5739 & 114 &  2 & 7.56 & 0.03 \\
 SiIV4089 &  80 &  2 & 7.54 & 0.05 \\
 SiIV4116 &  61 &  2 & 7.48 & 0.06 \\
%----------------------
\tableline
\noalign{\smallskip}
\multicolumn{3}{c}{} & 
%----------------------
 {\bf $\epsilon_{\rm Si}$\,=\, 7.53} & 
 \multicolumn{1}{r}{$\Delta\epsilon_{\rm Si}$($\sigma$)\,=\,0.06} \\
 &  \multicolumn{2}{r}{$\Delta$\micro(Si)\,=\,0.3}  & $\Rightarrow$ &
 \multicolumn{1}{r}{$\Delta\epsilon_{\rm Si}$(\micro)\,=\,0.03} \\
%----------------------
\end{tabular}
\end{minipage}  
\begin{minipage}[c]{0.45\textwidth}
\begin{tabular}{ccccc}
\noalign{\smallskip}
\tableline\tableline
\noalign{\smallskip}
%----------------------
HD\,36591 & 
\multicolumn{3}{c}{\Teff\,=\,27200 K, \grav\,=\,4.1 dex} & \micro(O)=4.5 \\
%----------------------
\tableline
\noalign{\smallskip}
%----------------------
  OII3945 &  69 &  2 & 8.80 & 0.04 \\
  OII3954 &  86 &  2 & 8.64 & 0.04 \\
  OII3982 &  62 &  2 & 8.64 & 0.04 \\
  OII4317 & 125 &  5 & 8.77 & 0.06 \\
  OII4319 & 114 &  2 & 8.76 & 0.03 \\
  OII4366 & 102 &  4 & 8.62 & 0.07 \\
  OII4414 & 144 &  2 & 8.61 & 0.03 \\
  OII4416 & 124 &  8 & 8.70 & 0.12 \\
  OII4452 &  53 &  3 & 8.59 & 0.07 \\
  OII4638 & 109 &  2 & 8.94 & 0.04 \\
  OII4641 & 145 &  2 & 8.84 & 0.03 \\
  OII4650 & 159 &  4 & 8.62 & 0.06 \\
  OII4661 & 113 &  2 & 8.83 & 0.03 \\
  OII4673 &  44 &  5 & 8.81 & 0.13 \\
  OII4676 &  97 &  4 & 8.70 & 0.07 \\
  OII4696 &  30 &  3 & 8.74 & 0.10 \\
  OII6641 &  48 &  2 & 8.90 & 0.04 \\
  OII6721 &  69 &  2 & 8.79 & 0.03 \\
  OII4072 & 113 &  7 & 8.65 & 0.13 \\
  OII4076 & 133 &  4 & 8.70 & 0.07 \\
  OII4078 &  56 &  3 & 8.77 & 0.08 \\
  OII4086 &  68 &  6 & 8.83 & 0.14 \\
  OII4891 &  34 &  3 & 8.67 & 0.09 \\
  OII4906 &  53 &  3 & 8.67 & 0.07 \\
  OII4941 &  48 &  2 & 8.54 & 0.04 \\
  OII4943 &  68 &  3 & 8.57 & 0.06 \\
  OII4956 &  24 &  3 & 8.77 & 0.11 \\
%----------------------
\tableline
\noalign{\smallskip}
 \multicolumn{3}{c}{} & 
%----------------------
 {\bf $\epsilon_{\rm O}$\,=\, 8.71} & 
 \multicolumn{1}{r}{$\Delta\epsilon_{\rm O}$($\sigma$)\,=\,0.10} \\
 &  \multicolumn{2}{r}{$\Delta$\micro(O)\,=\,0.3}  & $\Rightarrow$ &
 \multicolumn{1}{r}{$\Delta\epsilon_{\rm O}$(\micro)\,=\,0.04} \\
% \multicolumn{3}{r}{$\Delta$\Teff\,=\,500, $\Delta$\grav\,=\,0.1} & $\Rightarrow$ & 
% \multicolumn{1}{r}{$\Delta\epsilon_{\rm O}$(SP)\,=\,0.02} \\
%----------------------
\end{tabular}
\end{minipage}  
%% Any table notes must follow the \end{tabular} command.
%\tablecomments{\footnotesize }
\end{center}
}\end{table}
%
% -----------------------------------------------------------------------
% -----------------------------------------------------------------------

% -----------------------------------------------------------------------
% --------------------           HD36959         ------------------------
% -----------------------------------------------------------------------
%
\begin{table}[!h]
{\scriptsize
\begin{center}
%----------------------
\begin{minipage}[c]{0.45\textwidth}
\caption{\scriptsize Results from the abundance analysis of 
%----------------------
HD\,36959 (B1\,V) 
\label{res6}}
%----------------------
\begin{tabular}{ccccc}
\noalign{\smallskip}
\tableline\tableline
\noalign{\smallskip}
%----------------------
HD\,36959 & 
\multicolumn{3}{c}{\Teff\,=\,25800 K, \grav\,=\,4.2 dex} & \micro(Si)=0.0  \\
%----------------------
\tableline
\noalign{\smallskip}
Line & $EW$ & $\Delta$$EW$ & $\epsilon_{\rm Si}$ & $\Delta\epsilon_{\rm Si}$ \\
 & (m\AA) & (m\AA) & (dex) & (dex) \\
\tableline
\noalign{\smallskip}
%----------------------
 SiII4128 &  16 &  3 & 7.52 & 0.12 \\
 SiII4130 &  17 &  3 & 7.44 & 0.12 \\
SiIII4552 & 142 &  1 & 7.48 & 0.02 \\
SiIII4567 & 124 &  1 & 7.48 & 0.02 \\
SiIII4574 &  90 &  1 & 7.54 & 0.02 \\
SiIII5739 &  99 &  2 & 7.51 & 0.04 \\
 SiIV4089 &  45 &  1 & 7.45 & 0.04 \\
 SiIV4116 &  36 &  3 & 7.47 & 0.12 \\
%----------------------
\tableline
\noalign{\smallskip}
\multicolumn{3}{c}{} & 
%----------------------
 {\bf $\epsilon_{\rm Si}$\,=\, 7.49} & 
 \multicolumn{1}{r}{$\Delta\epsilon_{\rm Si}$($\sigma$)\,=\,0.03} \\
 &  \multicolumn{2}{r}{$\Delta$\micro(Si)\,=\,0.5}  & $\Rightarrow$ &
 \multicolumn{1}{r}{$\Delta\epsilon_{\rm Si}$(\micro)\,=\,0.07} \\
%----------------------
\end{tabular}
\end{minipage}  
\begin{minipage}[c]{0.45\textwidth}
\begin{tabular}{ccccc}
\noalign{\smallskip}
\tableline\tableline
\noalign{\smallskip}
%----------------------
HD\,36959 & 
\multicolumn{3}{c}{\Teff\,=\,25800 K, \grav\,=\,4.2 dex} & \micro(O)=2.1 \\
%----------------------
\tableline
\noalign{\smallskip}
%----------------------
  OII3945 &  55 &  2 & 8.78 & 0.05 \\
  OII3954 &  73 &  2 & 8.69 & 0.04 \\
  OII3982 &  50 &  2 & 8.63 & 0.05 \\
  OII4317 &  98 &  6 & 8.76 & 0.10 \\
  OII4319 &  86 &  3 & 8.69 & 0.06 \\
  OII4366 &  81 &  5 & 8.62 & 0.10 \\
  OII4414 & 116 &  2 & 8.64 & 0.03 \\
  OII4416 &  99 &  3 & 8.72 & 0.05 \\
  OII4452 &  45 &  2 & 8.62 & 0.05 \\
  OII4641 & 113 &  2 & 8.75 & 0.04 \\
  OII4650 & 133 &  5 & 8.65 & 0.08 \\
  OII4661 &  89 &  2 & 8.76 & 0.04 \\
  OII4673 &  35 &  2 & 8.72 & 0.07 \\
  OII4676 &  75 &  3 & 8.61 & 0.07 \\
  OII4696 &  25 &  2 & 8.69 & 0.08 \\
  OII6641 &  30 &  2 & 8.78 & 0.06 \\
  OII6721 &  48 &  2 & 8.75 & 0.04 \\
  OII4072 &  94 &  6 & 8.72 & 0.13 \\
  OII4076 & 108 &  6 & 8.73 & 0.12 \\
  OII4078 &  46 &  3 & 8.80 & 0.09 \\
  OII4086 &  48 &  3 & 8.70 & 0.09 \\
  OII4891 &  26 &  2 & 8.67 & 0.07 \\
  OII4906 &  40 &  2 & 8.67 & 0.06 \\
  OII4941 &  39 &  2 & 8.62 & 0.06 \\
  OII4943 &  55 &  2 & 8.66 & 0.05 \\
%----------------------
\tableline
\noalign{\smallskip}
 \multicolumn{3}{c}{} & 
%----------------------
 {\bf $\epsilon_{\rm O}$\,=\, 8.70} & 
 \multicolumn{1}{r}{$\Delta\epsilon_{\rm O}$($\sigma$)\,=\,0.02} \\
 &  \multicolumn{2}{r}{$\Delta$\micro(O)\,=\,0.4}  & $\Rightarrow$ &
 \multicolumn{1}{r}{$\Delta\epsilon_{\rm O}$(\micro)\,=\,0.06} \\
% \multicolumn{3}{r}{$\Delta$\Teff\,=\,500, $\Delta$\grav\,=\,0.1} & $\Rightarrow$ & 
% \multicolumn{1}{r}{$\Delta\epsilon_{\rm O}$(SP)\,=\,0.05} \\
%----------------------
\end{tabular}
\end{minipage}  
%% Any table notes must follow the \end{tabular} command.
%\tablecomments{\footnotesize }
\end{center}
}\end{table}
%
% -----------------------------------------------------------------------
% -----------------------------------------------------------------------

% -----------------------------------------------------------------------
% --------------------           HD37744         ------------------------
% -----------------------------------------------------------------------
%
\begin{table}[!h]
{\scriptsize
\begin{center}
%----------------------
\begin{minipage}[c]{0.45\textwidth}
\caption{\scriptsize Results from the abundance analysis of 
%----------------------
HD\,37744 (B1.5\,V) 
\label{res7}}
%----------------------
\begin{tabular}{ccccc}
\noalign{\smallskip}
\tableline\tableline
\noalign{\smallskip}
%----------------------
HD\,37744 & 
\multicolumn{3}{c}{\Teff\,=\,23800 K, \grav\,=\,4.1 dex} & \micro(Si)=0.5  \\
%----------------------
\tableline
\noalign{\smallskip}
Line & $EW$ & $\Delta$$EW$ & $\epsilon_{\rm Si}$ & $\Delta\epsilon_{\rm Si}$ \\
 & (m\AA) & (m\AA) & (dex) & (dex) \\
\tableline
\noalign{\smallskip}
%----------------------
 SiII3856 &  20 &  4 & 7.49 & 0.22 \\
 SiII6371 &  18 &  6 & 7.57 & 0.22 \\
SiIII4552 & 147 &  3 & 7.57 & 0.05 \\
SiIII4567 & 120 &  4 & 7.49 & 0.07 \\
SiIII4574 &  88 &  4 & 7.59 & 0.09 \\
SiIII5739 &  86 &  4 & 7.45 & 0.08 \\
 SiIV4116 &  23 &  4 & 7.56 & 0.26 \\
%----------------------
\tableline
\noalign{\smallskip}
\multicolumn{3}{c}{} & 
%----------------------
 {\bf $\epsilon_{\rm Si}$\,=\, 7.53} & 
 \multicolumn{1}{r}{$\Delta\epsilon_{\rm Si}$($\sigma$)\,=\,0.05} \\
 &  \multicolumn{2}{r}{$\Delta$\micro(Si)\,=\,0.5}  & $\Rightarrow$ &
 \multicolumn{1}{r}{$\Delta\epsilon_{\rm Si}$(\micro)\,=\,0.04} \\
%----------------------
\end{tabular}
\end{minipage}  
\begin{minipage}[c]{0.45\textwidth}
\begin{tabular}{ccccc}
\noalign{\smallskip}
\tableline\tableline
\noalign{\smallskip}
%----------------------
HD\,37744 & 
\multicolumn{3}{c}{\Teff\,=\,23800 K, \grav\,=\,4.1 dex} & \micro(O)=3.6 \\
%----------------------
\tableline
\noalign{\smallskip}
%----------------------
  OII3945 &  45 &  5 & 8.73 & 0.13 \\
  OII3954 &  66 &  7 & 8.72 & 0.15 \\
  OII4317 &  72 &  7 & 8.60 & 0.13 \\
  OII4319 &  73 &  8 & 8.67 & 0.16 \\
  OII4366 &  67 &  6 & 8.56 & 0.13 \\
  OII4414 & 100 & 14 & 8.62 & 0.23 \\
  OII4416 &  76 &  7 & 8.59 & 0.14 \\
  OII4452 &  32 &  6 & 8.51 & 0.20 \\
  OII4641 &  99 &  5 & 8.76 & 0.09 \\
  OII4650 & 123 &  8 & 8.71 & 0.13 \\
  OII4661 &  79 &  5 & 8.81 & 0.10 \\
  OII4673 &  25 &  5 & 8.67 & 0.20 \\
  OII4676 &  71 &  6 & 8.76 & 0.13 \\
  OII4696 &  20 &  5 & 8.73 & 0.23 \\
  OII6721 &  33 &  7 & 8.74 & 0.19 \\
  OII4069 & 136 & 13 & 8.83 & 0.18 \\
  OII4078 &  36 &  6 & 8.71 & 0.21 \\
  OII4906 &  32 &  4 & 8.69 & 0.13 \\
  OII4941 &  30 &  5 & 8.62 & 0.17 \\
  OII4943 &  45 &  5 & 8.69 & 0.13 \\
  OII4956 &  14 &  3 & 8.79 & 0.17 \\
%----------------------
\tableline
\noalign{\smallskip}
 \multicolumn{3}{c}{} & 
%----------------------
 {\bf $\epsilon_{\rm O}$\,=\, 8.70} & 
 \multicolumn{1}{r}{$\Delta\epsilon_{\rm O}$($\sigma$)\,=\,0.07} \\
 &  \multicolumn{2}{r}{$\Delta$\micro(O)\,=\,1.4}  & $\Rightarrow$ &
 \multicolumn{1}{r}{$\Delta\epsilon_{\rm O}$(\micro)\,=\,0.06} \\
% \multicolumn{3}{r}{$\Delta$\Teff\,=\,500, $\Delta$\grav\,=\,0.1} & $\Rightarrow$ & 
% \multicolumn{1}{r}{$\Delta\epsilon_{\rm O}$(SP)\,=\,0.09} \\
%----------------------
\end{tabular}
\end{minipage}  
%% Any table notes must follow the \end{tabular} command.
%\tablecomments{\footnotesize }
\end{center}
}\end{table}
%
% -----------------------------------------------------------------------
% -----------------------------------------------------------------------

% -----------------------------------------------------------------------
% --------------------           HD35299         ------------------------
% -----------------------------------------------------------------------
%
\begin{table}[!h]
{\scriptsize
\begin{center}
%----------------------
\begin{minipage}[c]{0.45\textwidth}
\caption{\scriptsize Results from the abundance analysis of 
%----------------------
HD\,35299 (B1.5\,V) 
\label{res8}}
%----------------------
\begin{tabular}{ccccc}
\noalign{\smallskip}
\tableline\tableline
\noalign{\smallskip}
%----------------------
HD\,35299 & 
\multicolumn{3}{c}{\Teff\,=\,23700 K, \grav\,=\,4.2 dex} & \micro(Si)=0.5  \\
%----------------------
\tableline
\noalign{\smallskip}
Line & $EW$ & $\Delta$$EW$ & $\epsilon_{\rm Si}$ & $\Delta\epsilon_{\rm Si}$ \\
 & (m\AA) & (m\AA) & (dex) & (dex) \\
\tableline
\noalign{\smallskip}
%----------------------
 SiII3856 &  27 &  5 & 7.65 & 0.24 \\
 SiII3862 &  15 &  2 & 7.44 & 0.13 \\
 SiII6371 &  23 &  4 & 7.64 & 0.13 \\
SiIII4552 & 133 &  5 & 7.45 & 0.08 \\
SiIII4567 & 110 &  2 & 7.40 & 0.04 \\
SiIII4574 &  79 &  2 & 7.50 & 0.05 \\
SiIII5739 &  81 &  2 & 7.47 & 0.04 \\
 SiIV4089 &  26 &  1 & 7.59 & 0.06 \\
 SiIV4116 &  18 &  2 & 7.48 & 0.15 \\
%----------------------
\tableline
\noalign{\smallskip}
\multicolumn{3}{c}{} & 
%----------------------
 {\bf $\epsilon_{\rm Si}$\,=\, 7.50} & 
 \multicolumn{1}{r}{$\Delta\epsilon_{\rm Si}$($\sigma$)\,=\,0.08} \\
 &  \multicolumn{2}{r}{$\Delta$\micro(Si)\,=\,0.5}  & $\Rightarrow$ &
 \multicolumn{1}{r}{$\Delta\epsilon_{\rm Si}$(\micro)\,=\,0.02} \\
%----------------------
\end{tabular}
\end{minipage}  
\begin{minipage}[c]{0.45\textwidth}
\begin{tabular}{ccccc}
\noalign{\smallskip}
\tableline\tableline
\noalign{\smallskip}
%----------------------
HD\,35299 & 
\multicolumn{3}{c}{\Teff\,=\,23700 K, \grav\,=\,4.2 dex} & \micro(O)=2.8 \\
%----------------------
\tableline
\noalign{\smallskip}
%----------------------
  OII3945 &  44 &  2 & 8.81 & 0.06 \\
  OII3954 &  59 &  2 & 8.72 & 0.05 \\
  OII3982 &  38 &  2 & 8.62 & 0.06 \\
  OII4317 &  78 &  5 & 8.80 & 0.09 \\
  OII4319 &  66 &  3 & 8.69 & 0.07 \\
  OII4366 &  61 &  3 & 8.59 & 0.07 \\
  OII4414 &  95 &  5 & 8.70 & 0.09 \\
  OII4416 &  80 &  2 & 8.77 & 0.04 \\
  OII4452 &  32 &  2 & 8.61 & 0.07 \\
  OII4641 &  91 &  5 & 8.79 & 0.10 \\
  OII4650 & 112 &  5 & 8.73 & 0.09 \\
  OII4661 &  68 &  3 & 8.77 & 0.07 \\
  OII4673 &  25 &  2 & 8.76 & 0.08 \\
  OII4676 &  59 &  2 & 8.68 & 0.05 \\
  OII4696 &  17 &  2 & 8.70 & 0.10 \\
  OII6641 &  18 &  2 & 8.82 & 0.09 \\
  OII6721 &  29 &  3 & 8.77 & 0.09 \\
  OII4072 &  82 &  3 & 8.81 & 0.07 \\
  OII4076 &  94 &  4 & 8.79 & 0.08 \\
  OII4078 &  35 &  2 & 8.79 & 0.07 \\
  OII4086 &  36 &  2 & 8.67 & 0.07 \\
  OII4891 &  16 &  3 & 8.64 & 0.16 \\
  OII4906 &  28 &  2 & 8.70 & 0.07 \\
  OII4941 &  25 &  2 & 8.59 & 0.08 \\
  OII4943 &  40 &  2 & 8.71 & 0.06 \\
  OII4956 &  12 &  2 & 8.78 & 0.13 \\
%----------------------
\tableline
\noalign{\smallskip}
 \multicolumn{3}{c}{} & 
%----------------------
 {\bf $\epsilon_{\rm O}$\,=\, 8.72} & 
 \multicolumn{1}{r}{$\Delta\epsilon_{\rm O}$($\sigma$)\,=\,0.07} \\
 &  \multicolumn{2}{r}{$\Delta$\micro(O)\,=\,0.6}  & $\Rightarrow$ &
 \multicolumn{1}{r}{$\Delta\epsilon_{\rm O}$(\micro)\,=\,0.03} \\
% \multicolumn{3}{r}{$\Delta$\Teff\,=\,500, $\Delta$\grav\,=\,0.1} & $\Rightarrow$ & 
% \multicolumn{1}{r}{$\Delta\epsilon_{\rm O}$(SP)\,=\,0.09} \\
%----------------------
\end{tabular}
\end{minipage}  
%% Any table notes must follow the \end{tabular} command.
%\tablecomments{\footnotesize }
\end{center}
}\end{table}
%
% -----------------------------------------------------------------------
% -----------------------------------------------------------------------

% -----------------------------------------------------------------------
% --------------------           HD36285         ------------------------
% -----------------------------------------------------------------------
%
\begin{table}[!h]
{\scriptsize
\begin{center}
%----------------------
\begin{minipage}[c]{0.45\textwidth}
\caption{\scriptsize Results from the abundance analysis of 
%----------------------
HD\,36285 (B2\,V) 
\label{res9}}
%----------------------
\begin{tabular}{ccccc}
\noalign{\smallskip}
\tableline\tableline
\noalign{\smallskip}
%----------------------
HD\,36285 & 
\multicolumn{3}{c}{\Teff\,=\,20600 K, \grav\,=\,4.0 dex} & \micro(Si)=1.7  \\
%----------------------
\tableline
\noalign{\smallskip}
Line & $EW$ & $\Delta$$EW$ & $\epsilon_{\rm Si}$ & $\Delta\epsilon_{\rm Si}$ \\
 & (m\AA) & (m\AA) & (dex) & (dex) \\
\tableline
\noalign{\smallskip}
%----------------------
 SiII3856 &  41 &  2 & 7.38 & 0.07 \\
 SiII3862 &  37 &  5 & 7.52 & 0.19 \\
 SiII6347 &  61 &  4 & 7.60 & 0.09 \\
 SiII6371 &  45 &  3 & 7.53 & 0.08 \\
SiIII4552 & 110 &  1 & 7.49 & 0.02 \\
SiIII4567 &  88 &  2 & 7.44 & 0.04 \\
SiIII4574 &  59 &  1 & 7.51 & 0.03 \\
SiIII5739 &  53 &  2 & 7.46 & 0.05 \\
%----------------------
\tableline
\noalign{\smallskip}
\multicolumn{3}{c}{} & 
%----------------------
 {\bf $\epsilon_{\rm Si}$\,=\, 7.49} & 
 \multicolumn{1}{r}{$\Delta\epsilon_{\rm Si}$($\sigma$)\,=\,0.06} \\
 &  \multicolumn{2}{r}{$\Delta$\micro(Si)\,=\,0.5}  & $\Rightarrow$ &
 \multicolumn{1}{r}{$\Delta\epsilon_{\rm Si}$(\micro)\,=\,0.05} \\
%----------------------
\end{tabular}
\end{minipage}  
\begin{minipage}[c]{0.45\textwidth}
\begin{tabular}{ccccc}
\noalign{\smallskip}
\tableline\tableline
\noalign{\smallskip}
%----------------------
HD\,36285 & 
\multicolumn{3}{c}{\Teff\,=\,20600 K, \grav\,=\,4.0 dex} & \micro(O)=5.5 \\
%----------------------
\tableline
\noalign{\smallskip}
%----------------------
  OII3945 &  26 &  3 & 8.79 & 0.12 \\
  OII3954 &  41 &  3 & 8.79 & 0.09 \\
  OII4317 &  46 &  5 & 8.74 & 0.12 \\
  OII4319 &  42 &  3 & 8.69 & 0.08 \\
  OII4366 &  42 &  4 & 8.65 & 0.11 \\
  OII4414 &  63 &  3 & 8.70 & 0.06 \\
  OII4416 &  50 &  3 & 8.74 & 0.07 \\
  OII4452 &  26 &  3 & 8.88 & 0.12 \\
  OII4638 &  46 &  3 & 8.94 & 0.08 \\
  OII4650 &  78 &  4 & 8.70 & 0.07 \\
  OII4661 &  46 &  2 & 8.80 & 0.05 \\
  OII4673 &  12 &  3 & 8.68 & 0.21 \\
  OII4676 &  36 &  3 & 8.66 & 0.09 \\
  OII6721 &  15 &  4 & 8.99 & 0.23 \\
  OII4069 &  81 &  6 & 8.86 & 0.10 \\
  OII4072 &  55 &  4 & 8.80 & 0.11 \\
  OII4078 &  24 &  4 & 8.91 & 0.19 \\
  OII4086 &  24 & 10 & 8.78 & 0.48 \\
  OII4906 &  16 &  3 & 8.85 & 0.17 \\
  OII4941 &  16 &  5 & 8.84 & 0.29 \\
  OII4943 &  25 &  3 & 8.91 & 0.12 \\
%----------------------
\tableline
\noalign{\smallskip}
 \multicolumn{3}{c}{} & 
%----------------------
 {\bf $\epsilon_{\rm O}$\,=\, 8.80} & 
 \multicolumn{1}{r}{$\Delta\epsilon_{\rm O}$($\sigma$)\,=\,0.10} \\
 &  \multicolumn{2}{r}{$\Delta$\micro(O)\,=\,1.5}  & $\Rightarrow$ &
 \multicolumn{1}{r}{$\Delta\epsilon_{\rm O}$(\micro)\,=\,0.06} \\
% \multicolumn{3}{r}{$\Delta$\Teff\,=\,500, $\Delta$\grav\,=\,0.1} & $\Rightarrow$ & 
% \multicolumn{1}{r}{$\Delta\epsilon_{\rm O}$(SP)\,=\,0.13} \\
%----------------------
\end{tabular}
\end{minipage}  
%% Any table notes must follow the \end{tabular} command.
%\tablecomments{\footnotesize }
\end{center}
}\end{table}
%
% -----------------------------------------------------------------------
% -----------------------------------------------------------------------

% -----------------------------------------------------------------------
% --------------------           HD35039         ------------------------
% -----------------------------------------------------------------------
%
\begin{table}[!h]
{\scriptsize
\begin{center}
%----------------------
\begin{minipage}[c]{0.45\textwidth}
\caption{\scriptsize Results from the abundance analysis of 
%----------------------
HD\,35039 (B2\,V) 
\label{res10}}
%----------------------
\begin{tabular}{ccccc}
\noalign{\smallskip}
\tableline\tableline
\noalign{\smallskip}
%----------------------
HD\,35039 & 
\multicolumn{3}{c}{\Teff\,=\,19800 K, \grav\,=\,3.7 dex} & \micro(Si)=3.3  \\
%----------------------
\tableline
\noalign{\smallskip}
Line & $EW$ & $\Delta$$EW$ & $\epsilon_{\rm Si}$ & $\Delta\epsilon_{\rm Si}$ \\
 & (m\AA) & (m\AA) & (dex) & (dex) \\
\tableline
\noalign{\smallskip}
%----------------------
 SiII3856 &  61 &  2 & 7.63 & 0.06 \\
 SiII3862 &  45 &  5 & 7.54 & 0.15 \\
 SiII6347 &  75 & 10 & 7.51 & 0.17 \\
 SiII6371 &  55 &  2 & 7.46 & 0.04 \\
SiIII4552 & 125 & 10 & 7.49 & 0.15 \\
SiIII4567 & 103 &  5 & 7.47 & 0.09 \\
SiIII4574 &  67 &  5 & 7.52 & 0.11 \\
SiIII5739 &  69 &  5 & 7.58 & 0.10 \\
%----------------------
\tableline
\noalign{\smallskip}
\multicolumn{3}{c}{} & 
%----------------------
 {\bf $\epsilon_{\rm Si}$\,=\, 7.52} & 
 \multicolumn{1}{r}{$\Delta\epsilon_{\rm Si}$($\sigma$)\,=\,0.06} \\
 &  \multicolumn{2}{r}{$\Delta$\micro(Si)\,=\,1.0}  & $\Rightarrow$ &
 \multicolumn{1}{r}{$\Delta\epsilon_{\rm Si}$(\micro)\,=\,0.08} \\
%----------------------
\end{tabular}
\end{minipage}  
\begin{minipage}[c]{0.45\textwidth}
\begin{tabular}{ccccc}
\noalign{\smallskip}
\tableline\tableline
\noalign{\smallskip}
%----------------------
HD\,35039 & 
\multicolumn{3}{c}{\Teff\,=\,19800 K, \grav\,=\,3.7 dex} & \micro(O)=5.3 \\
%----------------------
\tableline
\noalign{\smallskip}
%----------------------
  OII3945 &  31 &  2 & 8.91 & 0.07 \\
  OII3954 &  41 &  2 & 8.79 & 0.06 \\
  OII3982 &  25 &  2 & 8.69 & 0.08 \\
  OII4317 &  50 &  6 & 8.80 & 0.14 \\
  OII4319 &  45 &  4 & 8.74 & 0.11 \\
  OII4414 &  64 &  2 & 8.72 & 0.04 \\
  OII4416 &  54 &  4 & 8.82 & 0.10 \\
  OII4641 &  64 &  3 & 8.80 & 0.07 \\
  OII4650 &  82 &  2 & 8.74 & 0.04 \\
  OII4661 &  46 &  3 & 8.79 & 0.08 \\
  OII4673 &  15 &  4 & 8.79 & 0.25 \\
  OII4676 &  39 &  2 & 8.71 & 0.06 \\
  OII4069 &  80 &  3 & 8.88 & 0.05 \\
  OII4072 &  59 &  3 & 8.91 & 0.08 \\
  OII4078 &  21 &  4 & 8.84 & 0.21 \\
  OII4086 &  25 &  4 & 8.84 & 0.18 \\
  OII4941 &  12 &  3 & 8.68 & 0.21 \\
  OII4943 &  21 &  3 & 8.81 & 0.14 \\
%----------------------
\tableline
\noalign{\smallskip}
 \multicolumn{3}{c}{} & 
%----------------------
 {\bf $\epsilon_{\rm O}$\,=\, 8.79} & 
 \multicolumn{1}{r}{$\Delta\epsilon_{\rm O}$($\sigma$)\,=\,0.07} \\
 &  \multicolumn{2}{r}{$\Delta$\micro(O)\,=\,1.5}  & $\Rightarrow$ &
 \multicolumn{1}{r}{$\Delta\epsilon_{\rm O}$(\micro)\,=\,0.07} \\
% \multicolumn{3}{r}{$\Delta$\Teff\,=\,500, $\Delta$\grav\,=\,0.1} & $\Rightarrow$ & 
% \multicolumn{1}{r}{$\Delta\epsilon_{\rm O}$(SP)\,=\,0.15} \\
%----------------------
\end{tabular}
\end{minipage}  
%% Any table notes must follow the \end{tabular} command.
%\tablecomments{\footnotesize }
\end{center}
}\end{table}
%
% -----------------------------------------------------------------------
% -----------------------------------------------------------------------

% -----------------------------------------------------------------------
% --------------------           HD36629         ------------------------
% -----------------------------------------------------------------------
%
\begin{table}[!h]
{\scriptsize
\begin{center}
%----------------------
\begin{minipage}[c]{0.45\textwidth}
\caption{\scriptsize Results from the abundance analysis of 
%----------------------
HD\,36629 (B2\,V) 
\label{res11}}
%----------------------
\begin{tabular}{ccccc}
\noalign{\smallskip}
\tableline\tableline
\noalign{\smallskip}
%----------------------
HD\,36629 & 
\multicolumn{3}{c}{\Teff\,=\,20000 K, \grav\,=\,4.1 dex} & \micro(Si)=1.0  \\
%----------------------
\tableline
\noalign{\smallskip}
Line & $EW$ & $\Delta$$EW$ & $\epsilon_{\rm Si}$ & $\Delta\epsilon_{\rm Si}$ \\
 & (m\AA) & (m\AA) & (dex) & (dex) \\
\tableline
\noalign{\smallskip}
%----------------------
 SiII3856 &  52 &  2 & 7.58 & 0.07 \\
 SiII3862 &  42 &  2 & 7.55 & 0.08 \\
 SiII6347 &  65 &  3 & 7.62 & 0.07 \\
 SiII6371 &  48 &  2 & 7.53 & 0.06 \\
SiIII4552 &  90 &  1 & 7.52 & 0.02 \\
SiIII4567 &  73 &  1 & 7.49 & 0.02 \\
SiIII4574 &  46 &  1 & 7.51 & 0.03 \\
SiIII5739 &  42 &  2 & 7.53 & 0.06 \\
%----------------------
\tableline
\noalign{\smallskip}
\multicolumn{3}{c}{} & 
%----------------------
 {\bf $\epsilon_{\rm Si}$\,=\, 7.54} & 
 \multicolumn{1}{r}{$\Delta\epsilon_{\rm Si}$($\sigma$)\,=\,0.04} \\
 &  \multicolumn{2}{r}{$\Delta$\micro(Si)\,=\,0.5}  & $\Rightarrow$ &
 \multicolumn{1}{r}{$\Delta\epsilon_{\rm Si}$(\micro)\,=\,0.05} \\
%----------------------
\end{tabular}
\end{minipage}  
\begin{minipage}[c]{0.45\textwidth}
\begin{tabular}{ccccc}
\noalign{\smallskip}
\tableline\tableline
\noalign{\smallskip}
%----------------------
HD\,36629 & 
\multicolumn{3}{c}{\Teff\,=\,20000 K, \grav\,=\,4.1 dex} & \micro(O)=6.0 \\
%----------------------
\tableline
\noalign{\smallskip}
%----------------------
  OII3945 &  24 &  2 & 8.91 & 0.09 \\
  OII3954 &  28 &  2 & 8.68 & 0.07 \\
  OII4317 &  37 &  3 & 8.77 & 0.09 \\
  OII4319 &  30 &  2 & 8.63 & 0.07 \\
  OII4366 &  30 &  2 & 8.59 & 0.07 \\
  OII4414 &  50 &  2 & 8.69 & 0.05 \\
  OII4416 &  40 &  2 & 8.75 & 0.06 \\
  OII4452 &  16 &  1 & 8.74 & 0.06 \\
  OII4638 &  36 &  1 & 8.95 & 0.03 \\
  OII4641 &  41 &  1 & 8.63 & 0.03 \\
  OII4650 &  59 &  3 & 8.65 & 0.07 \\
  OII4661 &  32 &  2 & 8.72 & 0.07 \\
  OII4673 &  11 &  3 & 8.82 & 0.23 \\
  OII4676 &  27 &  2 & 8.65 & 0.07 \\
  OII4069 &  64 &  3 & 8.88 & 0.06 \\
  OII4072 &  45 &  2 & 8.83 & 0.06 \\
  OII4076 &  55 &  3 & 8.85 & 0.08 \\
  OII4078 &  15 &  2 & 8.79 & 0.13 \\
  OII4943 &  16 &  3 & 8.83 & 0.17 \\
%----------------------
\tableline
\noalign{\smallskip}
 \multicolumn{3}{c}{} & 
%----------------------
 {\bf $\epsilon_{\rm O}$\,=\, 8.76} & 
 \multicolumn{1}{r}{$\Delta\epsilon_{\rm O}$($\sigma$)\,=\,0.10} \\
 &  \multicolumn{2}{r}{$\Delta$\micro(O)\,=\,1.7}  & $\Rightarrow$ &
 \multicolumn{1}{r}{$\Delta\epsilon_{\rm O}$(\micro)\,=\,0.06} \\
% \multicolumn{3}{r}{$\Delta$\Teff\,=\,500, $\Delta$\grav\,=\,0.1} & $\Rightarrow$ & 
% \multicolumn{1}{r}{$\Delta\epsilon_{\rm O}$(SP)\,=\,0.14} \\
%----------------------
\end{tabular}
\end{minipage}  
%% Any table notes must follow the \end{tabular} command.
%\tablecomments{\footnotesize }
\end{center}
}\end{table}
%
% -----------------------------------------------------------------------
% -----------------------------------------------------------------------

% -----------------------------------------------------------------------
% --------------------           HD36430         ------------------------
% -----------------------------------------------------------------------
%
\begin{table}[!h]
{\scriptsize
\begin{center}
%----------------------
\begin{minipage}[c]{0.45\textwidth}
\caption{\scriptsize Results from the abundance analysis of 
%----------------------
HD\,36430 (B2\,V) 
\label{res12}}
%----------------------
\begin{tabular}{ccccc}
\noalign{\smallskip}
\tableline\tableline
\noalign{\smallskip}
%----------------------
HD\,36430 & 
\multicolumn{3}{c}{\Teff\,=\,18400 K, \grav\,=\,4.1 dex} & \micro(Si)=3.5  \\
%----------------------
\tableline
\noalign{\smallskip}
Line & $EW$ & $\Delta$$EW$ & $\epsilon_{\rm Si}$ & $\Delta\epsilon_{\rm Si}$ \\
 & (m\AA) & (m\AA) & (dex) & (dex) \\
\tableline
\noalign{\smallskip}
%----------------------
 SiII3862 &  61 &  5 & 7.43 & 0.13 \\
 SiII6347 & 100 & 10 & 7.57 & 0.16 \\
 SiII6371 &  74 &  6 & 7.49 & 0.12 \\
SiIII4552 &  77 &  3 & 7.55 & 0.06 \\
SiIII4567 &  53 &  2 & 7.40 & 0.05 \\
SiIII4574 &  28 &  5 & 7.37 & 0.19 \\
SiIII5739 &  25 &  2 & 7.47 & 0.08 \\
%----------------------
\tableline
\noalign{\smallskip}
\multicolumn{3}{c}{} & 
%----------------------
 {\bf $\epsilon_{\rm Si}$\,=\, 7.47} & 
 \multicolumn{1}{r}{$\Delta\epsilon_{\rm Si}$($\sigma$)\,=\,0.08} \\
 &  \multicolumn{2}{r}{$\Delta$\micro(Si)\,=\,1.0}  & $\Rightarrow$ &
 \multicolumn{1}{r}{$\Delta\epsilon_{\rm Si}$(\micro)\,=\,0.07} \\
%----------------------
\end{tabular}
\end{minipage}  
\begin{minipage}[c]{0.45\textwidth}
\begin{tabular}{ccccc}
\noalign{\smallskip}
\tableline\tableline
\noalign{\smallskip}
%----------------------
HD\,36430 & 
\multicolumn{3}{c}{\Teff\,=\,18400 K, \grav\,=\,4.1 dex} & \micro(O)=6.3 \\
%----------------------
\tableline
\noalign{\smallskip}
%----------------------
  OII3954 &  19 &  5 & 8.77 & 0.26 \\
  OII4317 &  20 &  5 & 8.75 & 0.24 \\
  OII4319 &  20 &  5 & 8.76 & 0.25 \\
  OII4366 &  18 &  4 & 8.64 & 0.21 \\
  OII4414 &  33 &  3 & 8.78 & 0.10 \\
  OII4416 &  26 &  7 & 8.85 & 0.29 \\
  OII4641 &  27 &  5 & 8.74 & 0.20 \\
  OII4650 &  36 &  2 & 8.66 & 0.06 \\
  OII4661 &  18 &  3 & 8.72 & 0.16 \\
  OII4676 &  18 &  4 & 8.78 & 0.21 \\
  OII4069 &  37 &  3 & 8.86 & 0.09 \\
  OII4072 &  27 &  4 & 8.82 & 0.18 \\
%----------------------
\tableline
\noalign{\smallskip}
 \multicolumn{3}{c}{} & 
%----------------------
 {\bf $\epsilon_{\rm O}$\,=\, 8.76} & 
 \multicolumn{1}{r}{$\Delta\epsilon_{\rm O}$($\sigma$)\,=\,0.07} \\
 &  \multicolumn{2}{r}{$\Delta$\micro(O)\,=\,2.2}  & $\Rightarrow$ &
 \multicolumn{1}{r}{$\Delta\epsilon_{\rm O}$(\micro)\,=\,0.08} \\
% \multicolumn{3}{r}{$\Delta$\Teff\,=\,500, $\Delta$\grav\,=\,0.1} & $\Rightarrow$ & 
% \multicolumn{1}{r}{$\Delta\epsilon_{\rm O}$(SP)\,=\,0.13} \\
%----------------------
\end{tabular}
\end{minipage}  
%% Any table notes must follow the \end{tabular} command.
%\tablecomments{\footnotesize }
\end{center}
}\end{table}
%
% -----------------------------------------------------------------------
% -----------------------------------------------------------------------

% -----------------------------------------------------------------------
% --------------------           HD35912         ------------------------
% -----------------------------------------------------------------------
%
\begin{table}[!h]
{\scriptsize
\begin{center}
%----------------------
\begin{minipage}[c]{0.45\textwidth}
\caption{\scriptsize Results from the abundance analysis of 
%----------------------
HD\,35912 (B2\,V) 
\label{res13}}
%----------------------
\begin{tabular}{ccccc}
\noalign{\smallskip}
\tableline\tableline
\noalign{\smallskip}
%----------------------
HD\,35912 & 
\multicolumn{3}{c}{\Teff\,=\,18500 K, \grav\,=\,4.0 dex} & \micro(Si)=3.2  \\
%----------------------
\tableline
\noalign{\smallskip}
Line & $EW$ & $\Delta$$EW$ & $\epsilon_{\rm Si}$ & $\Delta\epsilon_{\rm Si}$ \\
 & (m\AA) & (m\AA) & (dex) & (dex) \\
\tableline
\noalign{\smallskip}
%----------------------
 SiII3856 &  65 &  2 & 7.35 & 0.05 \\
 SiII3862 &  62 &  4 & 7.53 & 0.11 \\
 SiII6347 &  99 &  2 & 7.60 & 0.03 \\
 SiII6371 &  71 &  3 & 7.49 & 0.06 \\
SiIII4552 &  77 &  2 & 7.48 & 0.04 \\
SiIII4567 &  60 &  2 & 7.45 & 0.05 \\
SiIII4574 &  35 &  1 & 7.47 & 0.03 \\
SiIII5739 &  30 &  2 & 7.51 & 0.07 \\
%----------------------
\tableline
\noalign{\smallskip}
\multicolumn{3}{c}{} & 
%----------------------
 {\bf $\epsilon_{\rm Si}$\,=\, 7.48} & 
 \multicolumn{1}{r}{$\Delta\epsilon_{\rm Si}$($\sigma$)\,=\,0.07} \\
 &  \multicolumn{2}{r}{$\Delta$\micro(Si)\,=\,0.5}  & $\Rightarrow$ &
 \multicolumn{1}{r}{$\Delta\epsilon_{\rm Si}$(\micro)\,=\,0.04} \\
%----------------------
\end{tabular}
\end{minipage}  
\begin{minipage}[c]{0.45\textwidth}
\begin{tabular}{ccccc}
\noalign{\smallskip}
\tableline\tableline
\noalign{\smallskip}
%----------------------
HD\,35912 & 
\multicolumn{3}{c}{\Teff\,=\,18500 K, \grav\,=\,4.0 dex} & \micro(O)=6.3 \\
%----------------------
\tableline
\noalign{\smallskip}
%----------------------
  OII3945 &  15 &  5 & 8.88 & 0.32 \\
  OII3954 &  22 &  4 & 8.79 & 0.18 \\
  OII4317 &  25 &  5 & 8.82 & 0.20 \\
  OII4319 &  21 &  3 & 8.70 & 0.14 \\
  OII4366 &  22 &  5 & 8.70 & 0.23 \\
  OII4414 &  34 &  4 & 8.72 & 0.13 \\
  OII4416 &  29 &  3 & 8.85 & 0.11 \\
  OII4650 &  42 &  2 & 8.70 & 0.06 \\
  OII4661 &  25 &  3 & 8.86 & 0.13 \\
  OII4676 &  17 &  4 & 8.65 & 0.22 \\
  OII4069 &  44 &  3 & 8.91 & 0.08 \\
  OII4072 &  33 &  3 & 8.91 & 0.12 \\
%----------------------
\tableline
\noalign{\smallskip}
 \multicolumn{3}{c}{} & 
%----------------------
 {\bf $\epsilon_{\rm O}$\,=\, 8.79} & 
 \multicolumn{1}{r}{$\Delta\epsilon_{\rm O}$($\sigma$)\,=\,0.09} \\
 &  \multicolumn{2}{r}{$\Delta$\micro(O)\,=\,2.2}  & $\Rightarrow$ &
 \multicolumn{1}{r}{$\Delta\epsilon_{\rm O}$(\micro)\,=\,0.08} \\
% \multicolumn{3}{r}{$\Delta$\Teff\,=\,500, $\Delta$\grav\,=\,0.1} & $\Rightarrow$ & 
% \multicolumn{1}{r}{$\Delta\epsilon_{\rm O}$(SP)\,=\,0.13} \\
%----------------------
\end{tabular}
\end{minipage}  
%% Any table notes must follow the \end{tabular} command.
%\tablecomments{\footnotesize }
\end{center}
}\end{table}
%
% -----------------------------------------------------------------------
% -----------------------------------------------------------------------

.\\ \newpage 
.\\ \newpage 
.\\ \newpage 
.\\ \newpage 
.\\ \newpage 
.\\ \newpage 
.\\ \newpage 
.\\ \newpage 
.\\ \newpage 
.\\ \newpage 
.\\ \newpage 
.\\ \newpage 

%%%%%%%%%%%%%%%%%%%%%%%%%%%%%%%%%%%%%%%%%%%%%%%%%%%%%%%%%%%%%%%%%%%%%%%%%%%%
\begin{figure*}[!t]
\centering
\includegraphics[angle=180,scale=0.85]{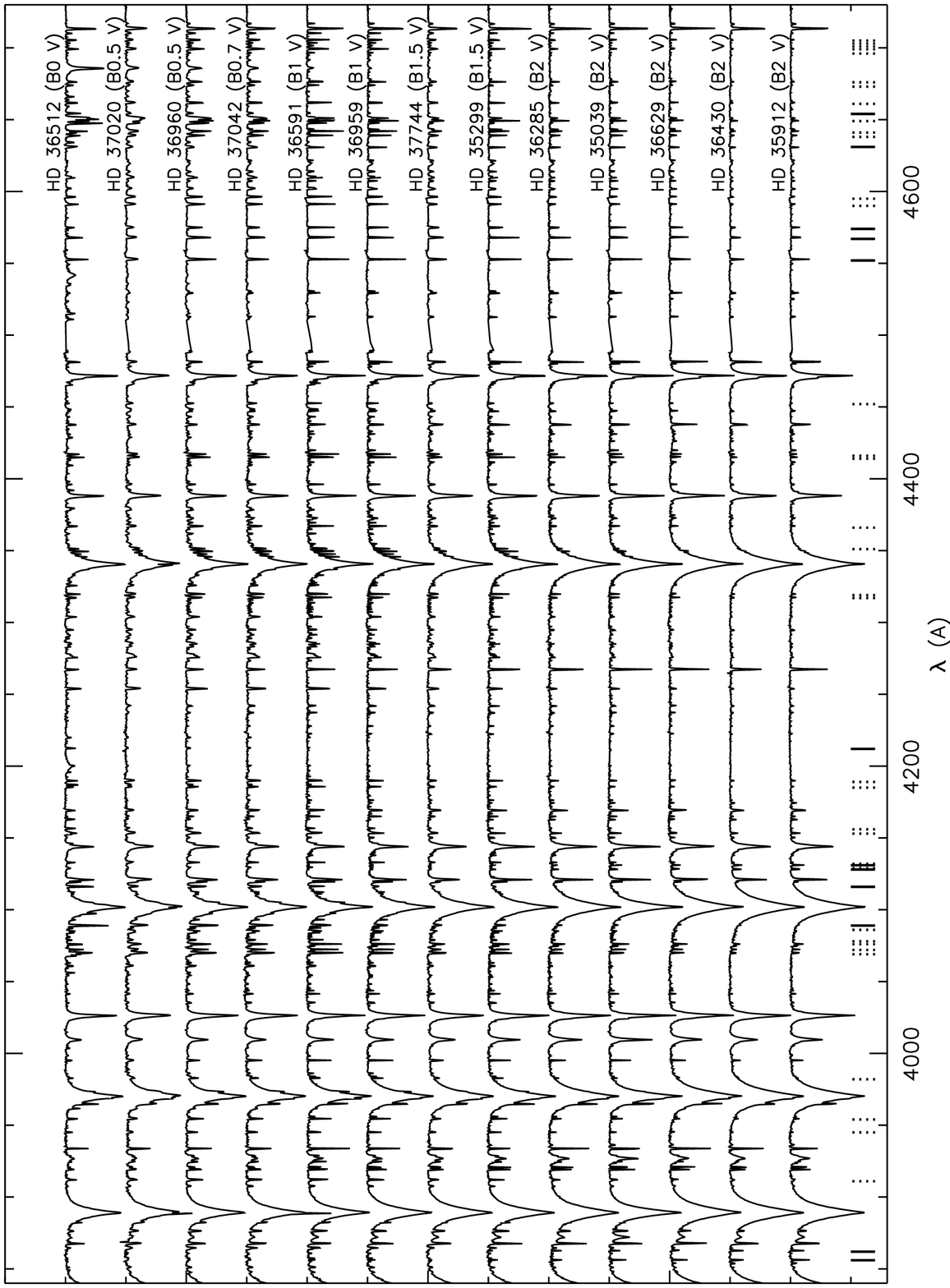}
\caption{The complete atlas of \fies\ spectra (part 1 of 3). The \ioni{Si}{ii-iv} and 
\ioni{O}{ii} lines used for the abundance analysis are indicated as solid 
and dashed vertical lines, respectively. \label{fig1e}}
\end{figure*}
%%%%%%%%%%%%%%%%%%%%%%%%%%%%%%%%%%%%%%%%%%%%%%%%%%%%%%%%%%%%%%%%%%%%%%%%%%%%

%%%%%%%%%%%%%%%%%%%%%%%%%%%%%%%%%%%%%%%%%%%%%%%%%%%%%%%%%%%%%%%%%%%%%%%%%%%%
\begin{figure*}[!t]
\centering
\includegraphics[angle=180,scale=0.85]{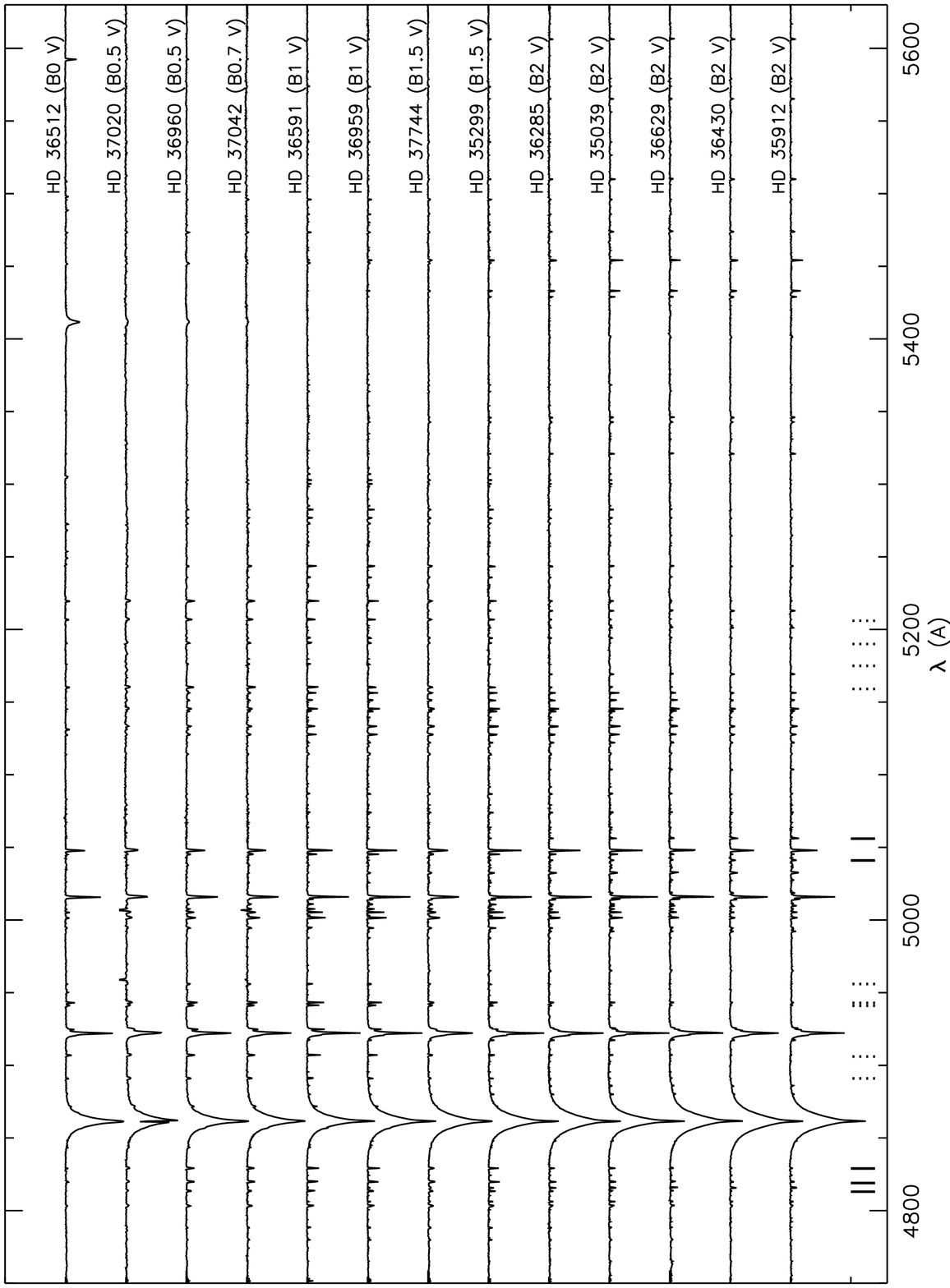}
\caption{The complete atlas of \fies\ spectra (part 2 of 3). The \ioni{Si}{ii-iv} and 
\ioni{O}{ii} lines used for the abundance analysis are indicated as solid 
and dashed vertical lines, respectively. \label{fig1e}}
\end{figure*}
%%%%%%%%%%%%%%%%%%%%%%%%%%%%%%%%%%%%%%%%%%%%%%%%%%%%%%%%%%%%%%%%%%%%%%%%%%%%

%%%%%%%%%%%%%%%%%%%%%%%%%%%%%%%%%%%%%%%%%%%%%%%%%%%%%%%%%%%%%%%%%%%%%%%%%%%%
\begin{figure*}[!t]
\centering
\includegraphics[angle=180,scale=0.85]{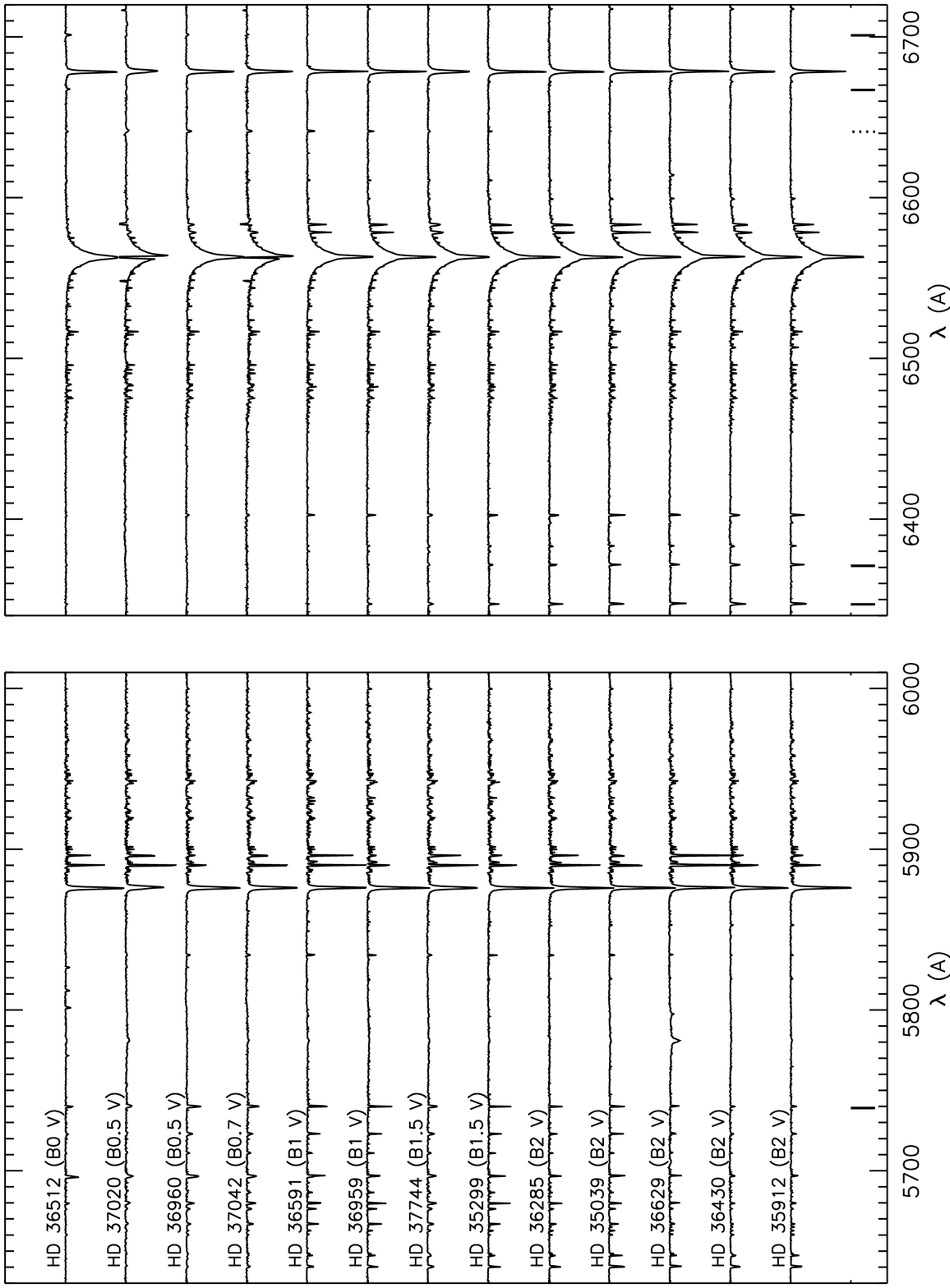}
\caption{The complete atlas of \fies\ spectra (part 3 of 3). The \ioni{Si}{ii-iv} and 
\ioni{O}{ii} lines used for the abundance analysis are indicated as solid 
and dashed vertical lines, respectively. \label{fig1e}}
\end{figure*}
%%%%%%%%%%%%%%%%%%%%%%%%%%%%%%%%%%%%%%%%%%%%%%%%%%%%%%%%%%%%%%%%%%%%%%%%%%%%

\end{document}